\documentclass[10pt,a4paper,final]{article}
\usepackage[utf8]{inputenc}
\usepackage{amsmath}
\usepackage{amsfonts}
\usepackage{amssymb}
\usepackage{graphicx}
\usepackage{hyperref}
\usepackage[left=2cm,right=2cm,top=2cm,bottom=2cm]{geometry}
\author{Henri Alloul \thanks{Henri Alloul (2015), NMR in strongly correlated materials, Scholarpedia, 10(1):30632, doi: \href{http://dx.DOI.org/10.4249/scholarpedia.30632}{10.4249/scholarpedia.30632} .}\\ LPS - CNRS/Universite Paris Sud, Orsay, France}
\title{{\bf NMR in strongly correlated materials}}
\date{}
\begin{document}
\maketitle
\begin{abstract}
Electron-electron interactions are at the origin of many exotic electronic properties of materials which have emerged from recent experimental observations. The main important phenomena discovered are related with electronic magnetic properties, which have been quite accessible to Nuclear Magnetic Resonance techniques. Those specifically permit to distinguish the orbitals or electronic bands responsible for the magnetism, the metallic properties and superconductivity and to reveal the physical properties which are distinct from expectations in an independent electron scheme. The description of some selected experimental cases permits us to underline the importance of the technique and to reveal altogether to the reader a wide range of novel phenomena specific to correlated electron physics. 
\end{abstract}

\section{Introduction}
\label{Intro}
In non interacting electronic systems, one considers energy levels with spin degeneracy and fills them with two electrons par level, without any consideration of U, the local coulomb repulsion on atomic orbitals. But as soon as one considers a solid which displays magnetic properties the latter has to be considered, as U is responsible for atomic and solid state magnetism. An introduction to these aspects has been given in a previous Scholarpedia article on the electronic properties of {\it Strongly Correlated Electron Systems}, which will be quoted as {\bf SCES} throughout this article.

If one starts with a completely free electron gas, the first incidence of weak correlations can be expressed in a Fermi liquid approach, that is the electronic states at the Fermi level are not single particle states but rather quasiparticle states in which the electron is dressed by an electronic cloud which involves the electronic correlations. Those quasiparticles are populated in a same way as free electron cases, except that the population jump at the Fermi level is smaller than unity. Correspondingly these quasiparticles have effective masses $m^\star$ which differ from the electron mass. This is seen for instance in the specific heat and the Pauli susceptibility.

With increasing electron correlations one reaches situations where electron states are in an intermediate regime between independent extended electronic states and local states. Those intermediate electronic states are at the basis of the correlated electron physics which gives exotic properties to the materials and various competing low $T$ states which are far from being understood at this time.

Here we shall take advantage of a series of NMR experimental investigations done on correlated electron systems, to introduce various specific effects which have been highlighted in such systems. The principal useful NMR parameters and technical details have been introduced in a previous Scholarpedia article {\it NMR studies of electronic properties of solids} that we shall quote as {\bf NMREPS} from now on. The good knowledge of the NMR characteristics in solids for which non interacting electron theories apply quite well, naturally permitted in the initial experiments to detect the unexpected modifications of electronic properties which occur in the presence of strong electronic correlations. This appears as an advantage of the NMR technique, with respect to most recent experimental probes which have been developed specifically to study strongly correlated electron systems.

This article will be organised as follows. We shall recall first in  {\bf section~\ref{Magnetic imputrities & Kundo effect}} the relatively simple case of the NMR studies on the magnetic properties of {\it 3d} impurities in metallic {\it sp} systems, which has been highlighted as the Kondo effect. This has been the earliest correlated electron physics case which has been understood. It has opened the way to the study of Heavy Fermions and Kondo lattices which will be touched in {\bf section~\ref{Heavy Fermions and Kondo Lattices}}.

The High $T_c$ cuprates is of course the family of compounds which has attracted a large interest on correlated electron physics especially in the low doping part of the phase diagram where NMR experiments have permitted to reveal the occurrence of a pseudogap as is detailed in {\bf section~\ref{Cuprate pseudogap}}. The original properties induced by electron interactions in 1D systems, that is Luttinger liquids will be briefly mentioned then in {\bf section~\ref{1D metals}}. We detail in {\bf section~\ref{Impurities and local responses in correlated electron systems}} the original behavior of impurities in correlated electron systems, spin chains, cuprates, which have been important to reveal some of the physical properties which were difficult to probe by distinct approaches. This altogether has induced large efforts to clarify the incidence of disorder on the properties of correlated electron systems. The study of exotic superconductivities and the capability of NMR to give some hints on the SC order parameter symmetry is illustrated in {\bf section~\ref{Exotic superconductivities}}. An important tendency towards charge ordering situations has been proposed to dominate the electronic properties of correlated electron systems. We shall illustrate in {\bf section~\ref{Charge differentiation and orbital order}}, in the particular case of Na cobaltates, that NMR ideally permits to unravel such situations in great detail. Of course purely magnetic insulating states are essential cases of correlated electron physics. Those imply a large variety of magnetic states, from ordered spin lattices to disordered spin glasses and spin liquid states which are highlighted when frustration effects occur in an ordered lattice. Some of the specific information which can be brought by NMR experiments on such magnetic states are discussed in {\bf section~\ref{Insulating Magnetic states and spin liquids}}. Finally we illustrate in {\bf section~\ref{Metallic magnetic states and Mott insulator to metal transition}} how NMR techniques permitted to study recently the insulator to metal transition induced by pressure in undoped half filled systems, that is the actual Mott transition. This has been made possible by the recent discovery of quasi 2D organic and 3D alkali fulleride compounds, which display quasi ideal 2D or 3D Mott transitions. Let us point out that throughout this article we restrict ourselves to a presentation of some robust experimental evidences on these correlated electron systems. We avoid as much as possible entering into the theoretical debates which are natural in a vivid research area and are not solved so far. 

\section{Magnetic impurities and Kondo effect}
\label{Magnetic imputrities & Kundo effect}
One of the first correlated electron physics problem which has been fully solved has been revealed by studies of $3d$ impurities substituted on the atomic sites of regular $sp$ metals. One usually assumed that a local moment $S$ resides on the $3d$ sites and interacts with the free electron spin $s$ by an exchange interaction
\begin{equation}
\label{exchange} 
H=-J\ S.\ s\ \delta (r)
\end{equation}
The Kondo problem arose with the discovery by J.Kondo that perturbation theory of this Hamiltonian resulted in a $-lnT$ term in the resistivity of the alloys, which was indeed observed experimentally. It was understood that the conduction electron interaction with the local moment induced a crossover of the impurity electronic state towards a low $T$ ground state quite different from the quasi-free local moment and that the crossover temperature defines an energy scale

\begin{equation}
\label{Kondo temp} 
k_{B}T_{K}=E_{F}\exp\left[ \frac{1}{J\rho (E_{F})}\right]
\end{equation}

This expression for the Kondo temperature $T_{K}$ bears some analogy with that of $T_{c}$ and the energy gap variation with electron-phonon coupling for superconductivity. It has been harder to qualify the actual properties of the Kondo ground state, but from the observed transport and thermodynamic properties associated with the impurity degrees of freedom, it has been accepted rather soon that the impurity properties experimentally appear to evolve from a high $T$ magnetic state to a non-magnetic like behavior below $T_{K}$. In other words, the weak coupling regime where the impurity moment can be treated in a perturbation scheme evolves at low $T$ in a strong coupling regime where the impurity and conduction electrons are bound into the ground state. The basic picture which was initially accepted is that the conduction electrons might form a singlet state with the impurity and compensate its magnetization. If such a spatially extended state occurs, one would then expect to see its experimental signature on local magnetic measurements in the corresponding spatial range around the impurity, so that NMR experiments appeared as the ideal probe to view such effects.From the study of the macroscopic properties of impurities in noble metal hosts, it was established that the crossover temperature $T_{K}$ was highly
dependent on the impurity. This was of course quite compatible with the
exponential expression of Eq.2. Values of $T_{K}$ could be estimated from the maximum in the impurity contribution to the specific heat, or from the Weiss contribution to the spin susceptibility measured at high enough temperature, etc. This permitted to establish that $T_{K}$ was below 10 mK for Cu-Mn, $\sim1$ K for Cu-Cr, $\sim30$ K for Cu-Fe, $\sim300$ K for Au-V, etc. (Daybell and Steyert 1968).It was harder to consider Al-Mn along the same lines as all temperature variations were very weak in this case,so that this crossover could only occur above 1000 K, for which the alloy would have molten. Anyway, if one wanted to study experimentally the change from the magnetic state to the non magnetic state, one needed to consider in priority a system in which one can explore both regimes $T>>T_{K}$ and $T<<T_{K}$. Therefore Cu-Fe appeared immediately as the most suitable case if one wanted to avoid extremely low temperature experiments, while Cu-Mn and Al-Mn appeared as the two extreme opposite cases.

\subsection{Spatial extent of the Kondo singlet and T dependence of the susceptibility}

This idea of a Kondo singlet has led to some attempts to detect
modifications of the host $^{63}$Cu NMR width when $T$ is decreased through $T_{K}$.Those early experiments were initially taken as a signature of the development of a static polarized cloud anti-parallel to the local impurity magnetization. But the situation was only fully clarified when NMR resonances of $^{63}$Cu near neighbors to the Fe were detected (see {\bf NMREPS}). The shifts of the various lines had $T$ variations which scaled with each-other and displayed the same Curie-Weiss dependence as the magnetic susceptibility data taken in very dilute samples, as displayed in {\bf Fig.~\ref{Fig1}(a)}. So, on a small number of sites near the impurity, the magnetic behaviour does display a smooth $T$ variation through $T_{K}$, which allowed one to deny the existence of a static compensating cloud.This result confirmed that the susceptibility reaches a low $T$ behavior similar to that achieved in a non-magnetic case, as has been also found by the numerical solutions of the Kondo model established by Wilson (Nobel prize) (Wilson 1975). However, these results do not give any answer about the spatial
extension of the correlated Kondo state (this matter is discussed in
Alloul, 2012).

\subsection{The electronic dynamic response}
\begin{figure}
\centering
\includegraphics[height=8cm,width=10cm]{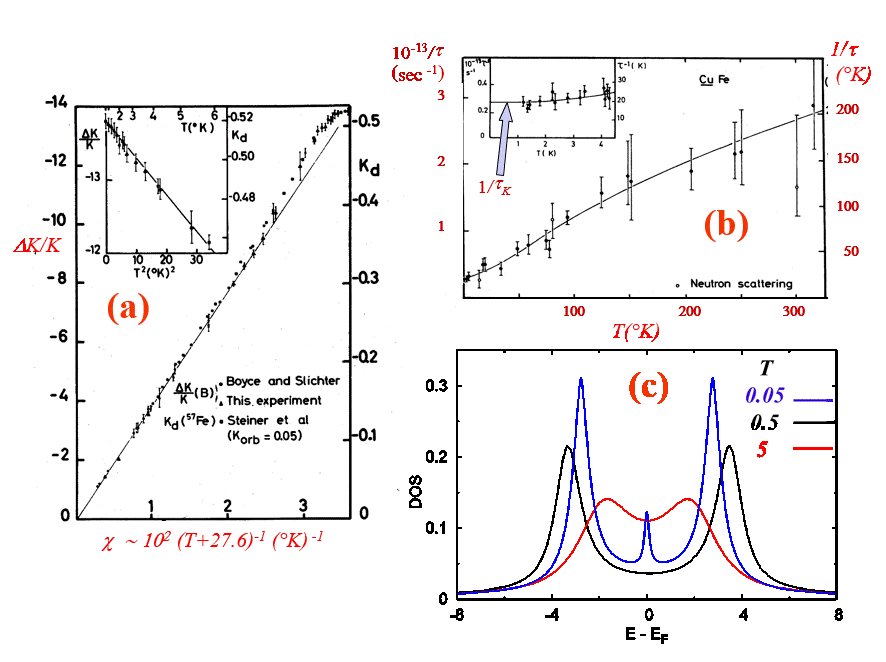} 
\caption{{\bf(a)} The variations of the normalized NMR shift $\Delta K/K$ induced by Fe impurities in Cu for three satellite resonances, and of the impurity susceptibility obtained by Mössbauer effect data ($K_{d}$)
scale perfectly with each-other. This gives a good experimental
determination of the variation of the local spin susceptibility through $T_{K}\sim30$K, which crosses over from a high $T$ Curie-Weiss
dependence above $T_{K}\ $toward a quadratic $T^{2}$ variation below $T_{K}$ (see inset) (from Alloul, 1975 ). {\bf(b)} Experimental variation of $\tau ^{-1}$ deduced from the $(T_{1}T)^{-1}$ data for various near-neighbour sites of Fe in copper. Three data points obtained from neutron scattering experiments are plotted with empty symbols. The inset gives an expansion for the low $T$ scale values (adapted from Alloul 1976). {\bf (c)}  The computed spectral functions (from Janis and Augustinsky 2008) are plotted for different temperatures $T$. The spectral function for $T<<T_{K}$ give evidence for the two Hubbard bands and the Kondo resonance at the Fermi energy. The latter disappears at high temperature. In this figure all energies are normalized to the hybridization energy of the Anderson impurity model.}
\label{Fig1}
\end{figure}

A large effort has been done initially using ESR spectroscopy to try to
understand the dynamic properties of local moment impurities in metals. It is well known ({\bf NMREPS}) that nuclear spins $I$, which couple to the band $s$ electron spins in a metal by a scalar contact hyperfine coupling display a $T$ linear spin lattice relaxation rate $(T_{1})^{-1}$ . The local moment $S$ being coupled to the band electron spins $s$ by the analogouscontact Hamiltonian of Eq.(1) one then expects to obtain a similar Korringa local moment relaxation rate
\begin{equation}
\label{Fluct.K}
\tau^{-1}=(\frac{\pi}{\hbar })\ k_{B}T\ J^{2}\rho ^{2}(E_{F})
\end{equation}
Some technical complications limit the use of Electron Paramagnetic
Resonance (EPR). This has stimulated the use of nuclear spins as probes of the dynamics of the local moment. Indeed the occurrence of the spin density oscillations around the impurities, recalled in {\bf NMREPS} resumes in an indirect coupling between the nuclear spin $I_{n}\ $at site $R_{n}$ and the local moment at the origin, given by
\begin{equation}
\label{Transferred}H_{IS}=A(R_{n})I_{n}S
\end{equation}
It can be easily seen that the fluctuations of the local moment induce local field fluctuations on $I_{n}$ that are responsible for its nuclear spin lattice relaxation rate. The latter can be written as

\begin{equation}
\label{NuclearspinT1}
(1/T_{1})_{n}=2\hbar ^{-2}k_{B}T\ A^{2}(R_{n})\frac{\chi }{(h\gamma _{e})^{2}}\frac{\tau }{1+(\omega _{e}\tau )^{2}}
\end{equation}

assuming a standard Lorentzian shape for the local moment dynamic
susceptibility. Quantitative analyses could be obtained by $T_{1}$
measurements taken on the NMR signals of the individual near-neighbor
shells of the impurity. Combining the $T_{1}$ data with the measure of $
\Delta K(R_{n})$, the contribution of the local moment to the shift of given satellites permits to eliminate $A^{2}(R_{n}).$ The obtained values of $\tau^{-1}$ were indeed found independent of the observed satellite, and were furthermore $T$ independent below $T_{K}$ in Cu-Fe, as can be seen in {\bf Fig.~\ref{Fig1}(b)}. In the low $T$ limit, the data resumes into a Korringa relation 
\begin{equation}
\label{Korringa } 
T_{1}T\Delta K^{2}=\frac{\hbar }{4\pi k_{B}}\left( \frac{\gamma _{e}}{\gamma
_{n}}\right) ^{2}
\end{equation}
where $\gamma _{e}$ and $\gamma _{n}$ are the electron and nuclear
gyromagnetic ratios.
These experiments therefore established that $\tau ^{-1}$ reaches a
constant value in the Kondo state, as does the local moment susceptibility. This also means that the Kondo energy scale $k_{B}T_{K}$ which limits the increase of $\chi$ below $T_{K}$, governs as well the $T$ independent spin lattice rate $\tau ^{-1}$ for the local moment, with both $\chi\sim\mu _{B}^{2}\ /\ (k_{B}T_{K})$ and$\ h/\tau \sim
k_{B}T_{K}$ (Alloul, 1977). Such a result then fits with the idea that the Kondo state has a dynamic behavior controlled by $k_{B}T_{K}$.
Let us point out that calculations done along the Renormalization Group Analysis or the Bethe Ansatz showed that the Kondo state displays a resonance at the Fermi energy (see {\bf Fig.~\ref{Fig1}(c)}), which
has a very narrow width controlled by$\ k_{B}T_{K}$, and that the local
moment spin lifetime appears as a direct determination of the width of this Kondo resonance peak.  Similar measurements of $1/\tau $ in Cu-Mn could be combined with the data taken in Cu-Fe to give the overall variation of $1/\tau $ over a few decades in $T/T_{K}$ extending below and above $T_{K}$ (Alloul {\it et al}, 1979).This permits to underline a transition from local moment physics at high $T$ to a Fermi liquid type of behaviour which can be considered as a local Fermi liquid.\\

{\bf In this section we have shown that a magnetic impurity substituted in a host metal induces a spin polarisation in its surroundings which can be probed by the NMR shift of the neighbouring host nuclei. This permits directly to monitor the $T$ variation of the impurity moment susceptibility and spin dynamics, the latter through $T_{1}$ measurements. Both quantities evolve from a local moment to a local Fermi liquid behaviour through a temperature $T_{K}$ determined by the Kondo energy scale of the impurity.}

\section{Heavy Fermions and Kondo Lattices}
\label{Heavy Fermions and Kondo Lattices}

\begin{figure}
\centering
\includegraphics[height=8cm,width=10cm]{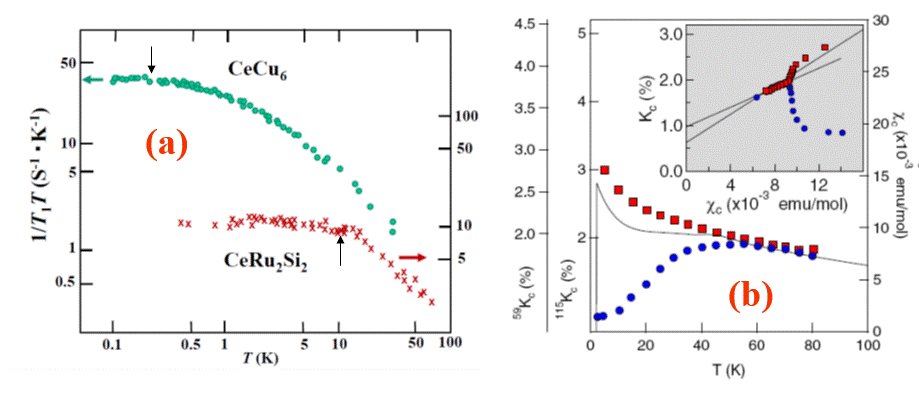}
\caption{{\bf(a)} $(T_{1}T)^{-1}$ is plotted versus temperature for $^{63}$Cu in CeCu$_{6}$ and $^{29}$Si in CeRu$_{2}$Si$_{2}$ (from Kitaoka {\it et al} 1987). These two compounds have respective Kondo temperatures $T_{K}$ of about 6 and 12 K. The arrows indicate the temperatures $T_{coh}$ below which the Korringa behaviour is reached that is respectively 0.2 and 8K. {\bf(b)} NMR shift anomaly in CeCoIn$_{5}$ which permits to locate $T_{coh}$ at about 50K in this compound. This can be seen from the non linear scaling of the In (red squares) and Co (blue circles) NMR shifts with the susceptibility (from Curro 2009).}
\label{Fig2}
\end{figure}

The Kondo effect discussed in {\bf section~\ref{Magnetic imputrities & Kundo effect}} applies as well for dilute rare earth ($4f$) or actinides ($5f$) impurities in metals. This has been found most often for Ce, Yb or U impurities for which the Kondo state resumes also into a narrow resonance at the Fermi level of the host material. The significant difference with $3d$ transition elements is that a quite large concentration of magnetic sites can be usually introduced in those intermetallic compounds. For instance, in
systems such as Ce$_{x}$La$_{1-x}$Cu$_{6}$ or Ce$_{x}$La$_{1-x}$B$_{6}$ the concentration of Ce can even reach $x=1$. In such compounds the behaviour of Ce can be considered as that of independent entities as long as the spin susceptibility and specific heat are roughly proportional to the Ce concentration.

\subsection{Local moment versus extended behaviour}
Among such families of quasi stoichiometric compounds, a recently
discovered one is that of CeCoIn$_{5}$ with one local moment per lattice
unit cell, for which the phase diagram with varying pressure reveals a
competition between magnetism and SC as in many other correlated electrons systems. In such stoichiometric compounds the strong electronic correlations are dominant and, if Kondo physics applies one deals then with a Kondo lattice of rare earth $f$ electron systems (see Kuramoto and Kitaoka, 2000). Interesting new phenomena occur in such dense systems as the moments of the $f$ electrons usually interact with a conduction band formed by $d$ electrons of the same or distinct atomic species. This raised questions about the possibility of Kondo singlet formation when the number of electrons available for Kondo screening is the same as that to be screened. In such {\bf Kondo lattices}, can
one still consider that the Kondo effect is similar to that of isolated
impurities? 

Of course, whether we consider the high $T$ local moment limit, or the low $T$ singlet state, the large concentration of sites leads to significant interactions that have to be considered. For the low $T$ Kondo singlet ground state, intersite interactions yield a metallic behavior for the $f$ electrons. In that case the Kondo state can be mapped out with very narrow bands with large DOS that one qualifies as heavy Fermions by reference to heavy Fermi liquids. Indeed the specific heat can be enhanced by orders of magnitude with respect to that expected from the bandwidth given by band structure calculations. The occurrence of these heavy quasiparticles has been seen by quantum oscillations detected by de Haas-van Alphen effect on the susceptibility in compounds such as UPt$_{3}$, CeRu$_{2}$Si$_{2}$ and CeCu$_{6}$(Julian {\it et al.}, 1994). These renormalized quasiparticles are of course expected to be at the origin of the rich variety of ground states which occur there, as in other correlated electron systems.
  
But what happens if one considers then the high $T$ local moment regime
above the Kondo temperature $T_{K}$ of the isolated impurities? There the nuclear spin lattice relaxation $1/T_{1}$ is nearly $T$ independent, and it only acquires a $T$ variation close to the Kondo temperature $T_{K}$ which is extracted from the analyses of the resistivity and the magnetic specific heat. As can be seen in {\bf Fig.~\ref{Fig2}(a)}, a Korringa like $T$ variation characteristic of the Fermi liquid ground state is only reached at much lower $T$. So, contrary to the single Kondo impurity case where $T_{K} $ is the single energy scale required to describe the physics, here a second energy scale is needed. It points out a regime often labelled as $T_{coh}$, where the Kondo singlets acquire a coherent band behavior (Yang Y. {\it et al.}, 2008). As displayed in {\bf Fig.~\ref{Fig2}(a)}, ratios $T_{coh}/T_{K}$ of 0.03 and 0.67 were respectively found for CeCu$_{6}$ and CeRu$_{2}$Si$_{2}$, which emphasizes that the two energy scales are material dependent. 
The existence of this coherence regime has been detected as well as a non linear scaling of the NMR shift with the susceptibility (Curro 2009), suggesting a change of hyperfine coupling with temperature as highlighted in {\bf Fig.~\ref{Fig2}(b)}. The existence of $T_{coh}$ is detected as well in the transport properties. Overall these results demonstrate that the heavy-electron states display both high $T$ local moment and low $T$ extended band behaviors.

\subsection{Heavy-Electron Magnetism}
\begin{figure}
\centering
\includegraphics[height=8cm,width=10cm]{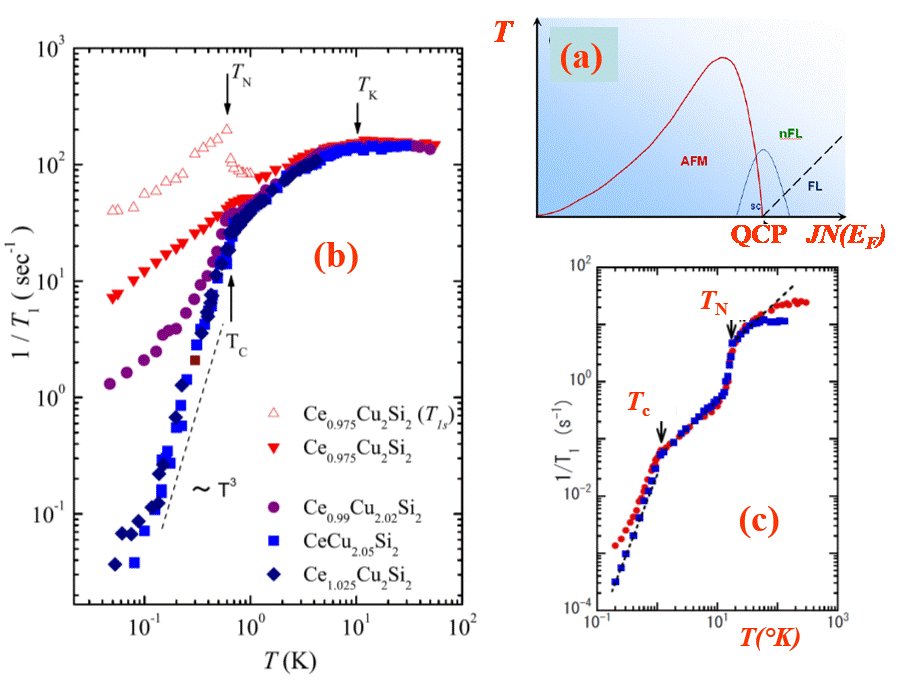}
\caption{{\bf(a)} Schematic phase diagram showing the variation of ordering temperature $T_{M}$ with increasing density of states. The ordered magnetism disappears at the QCP where usually underlying SC phases appear (from Curro 2009).{\bf(b)} In CeCu$_{2}$Si$_{2}$ the $T_{1}$ measured by NQR of the Cu site reveal that the SC state at $T_{c}\sim1$K appears below the Kondo temperature $T_{K}\sim10$K. In presence of impurities a local magnetic order is detected from the appearance of two components in the $T_{1}$ data (red symbols). The SC state appears below  $T_{N}$, suggesting a proximity to a {\bf QCP} (from Ishida {\it et al} 1999 ). {\bf (c)} In URu$_{2}$Si$_{2}$ superconductivity is seen to occur at $T_{c}\sim1$ K, within the magnetically ordered state which sets in at $T_{N}\sim 10$ K (from Matsuda {\it et al} 1996).}
\label{Fig3}
\end{figure}
So far we did not consider the Kondo lattice in the local moment regime, for which a spin polarization, mediated by the conduction electrons, is induced on its neighbors by a local moment (RKKY interaction - see {\bf NMREPS}). This induces an interaction between local moments which can be large enough to drive a magnetic order of the local moments at a temperature $k_{B}T_{M}$. Its energy scale is expected to be given by $\sim J_{int}\sim J^{2}`\rho(E_{F})$, where $J$ is the exchange interaction between $f$ and conduction electrons and $\rho(E_{F})$ is the density of conduction band states per site. But the Kondo effect which drives non-magnetic singlet ground states competes with the magnetic ordering tendency and might yield a reduction of $T_{M}$. For strong Kondo coupling such as $k_{B}T_{K}\sim J^{2}\rho(E_{F})$, any static
magnetic order should even be suppressed, as illustrated in the phase diagram of {\bf Fig.~\ref{Fig3}(a)} proposed initially by S. Doniach.

The main point which we shall underline hereafter is that depending of the system parameters one will get very different behaviors. By varying the magnitude of the density of states either by impurity substitutions or by applying a pressure, one could even induce a $T=0$ transition in a given system from a magnetic to a non magnetic state. Such a situation which is commonly underlined as a Quantum Critical Point ({\bf QCP}) is quite often encountered in correlated electron systems. We illustrate for instance in {\bf Fig.~\ref{Fig3}(b)} and {\bf (c)} two experimental situations respectively on the right and the left of a possible {\bf QCP}. In  {\bf Fig.~\ref{Fig3}(b)} one has a SC state below the Kondo temperature, while in {\bf Fig.~\ref{Fig3}(c)} a magnetic ordering, although with a very small moment (~$0.04\mu_{B}$ per site), occurs above the SC state.
 
This simple model describes some of the essential physics, but fails to
capture many details particularly in the vicinity of the {\bf QCP}, where superconductivity often emerges and where the behavior of the
normal state is usually not Fermi liquid like. Despite many years
of intensive theoretical work, there still remains no complete theory for the behavior of these heavy fermions in the vicinity of the {\bf QCP}. Clearly, in these cases the heavy quasiparticles do
form the superconducting condensate, and the hyperfine coupling to
these heavy quasiparticles is modified by the onset of coherence.\\

{\bf Some rare earth or uranide atoms can be involved in high concentration up to perfect stoichiometry in intermetallic compounds. The electronic state of their $f$ shells  displays a local moment behaviour at high $T$ but usually evolves below a coherence temperature towards a low $T$ metallic heavy Fermion state. Depending upon the magnitude of the interatomic interactions a magnetic ordered state of the atomic moments or a low $T$ Kondo lattice state can be favored. A controlled change of the system parameters might induce an evolution between these two states through a quantum critical point (QCP).}

\section{The Cuprate pseudogap}
\label{Cuprate pseudogap}
\begin{figure}
\centering
\includegraphics[height=8cm,width=10cm]{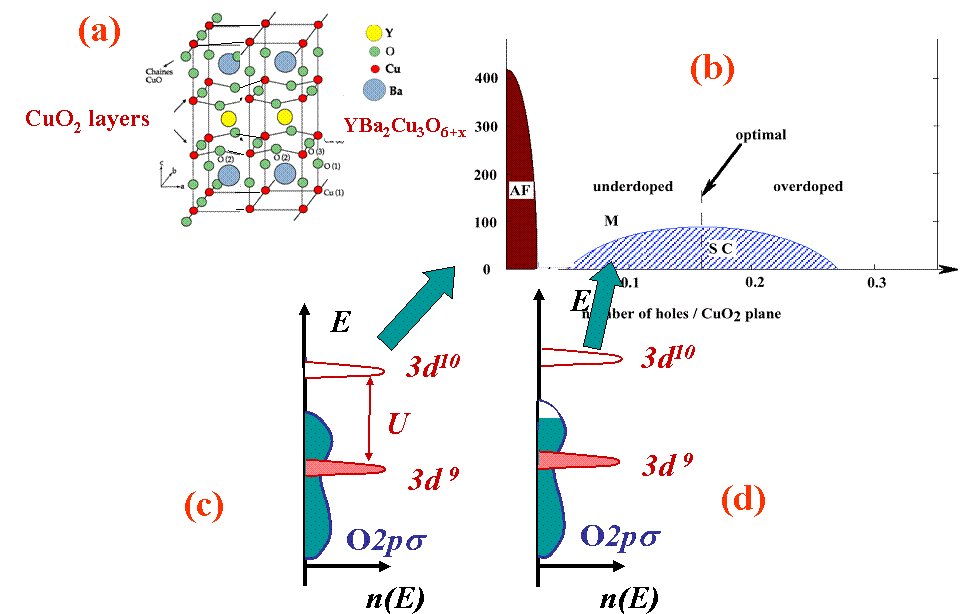}
\caption{{\bf(a)} Atomic structure of YBa$_{2}$Cu$_{3}$O$_{6+x}\ $.
The O filling of the lower and upper CuO$_{x}$ planes is
responsible for hole doping. {\bf (b)} Phase diagram versus hole doping
of the CuO$_{2}$ planes in cuprates showing the fast decrease of AF Néel
temperature, and the SC dome.{\bf (c)} Band structure of the undoped
parent compound. {\bf (d)} possible two band electronic structure in the
absence of interaction between Cu and O holes.}
\label{Fig4}
\end{figure}
The cuprates contain as common elements CuO$_{2}$ planes ({\bf Fig.~\ref{Fig4}(a)}) which are considered to contain all the important physics. Their structure is a stacking of such planes separated by other oxide layers, which maintain charge neutrality and the cohesion of the structure essentially through ionic interactions. They display the highest known superconducting temperature $T_{c}$ obtained after chemical doping a parent state which is a Mott insulator (see {\bf SCES}). Indeed in the undoped cuprates the Cu are in a $3d^{9}$ state in which the Cu hole bears a $S=1/2$ local moment({\bf Fig.~\ref{Fig4}(c)}). We shall discuss later on in {\bf section~\ref{Charge differentiation and orbital order}} how the magnetism of this Mott insulating state could be detected in NMR experiments. Those have been of course the physical properties which have driven the interest for these systems initially both for their fundamental aspects and their potential applications. Another aspect responsible for their appeal has certainly been the fact that the carrier concentration can be easily changed by chemical heterovalent substitutions or oxygen insertion in the layers separating the CuO$_{2}$ planes, which play the important role of charge reservoirs. Electron or hole doping can then be fairly continuously controlled from zero to about 0.3 charges per unit cell, which allows one to study the evolution of the physical properties of these materials with doping and to map out their rich phase diagram ({\bf Fig.~\ref{Fig4}(b)}). One important question raised concerning these doped compounds was that of the electronic structure of the band responsible for the metallic behavior. At a time where no ARPES experiments were available to map out the electronic structure, one expected that the doping holes would be located in an independent oxygen band, as exhibited in {\bf Fig.~\ref{Fig4}(d)}. As recalled hereafter in {\bf section~\ref{Carriers and Hyperfine couplings}}, the analysis of the $^{17}$O,$^{89}$Y and $^{63}$Cu NMR shifts in YBa$_{2}$Cu$_{3}$O$_{6+x}$ cuprates has permitted to demonstrate unambiguously that the holes responsible for the measured
macroscopic spin susceptibility are located on the Cu sites as expected for the undoped compound. The study of the evolution of the NMR shifts with hole doping ({\bf section~\ref{Evidence for a pseudogap from NMR shift data}}) allowed then to reveal the occurrence of a pseudogap in the samples with lower than optimal doping. The latter was found quite generic of the clean cuprate families ({\bf section~\ref{Universality of the pseudogap phase diagram}}). The analysis of the NMR spin lattice relaxation suggested a {\bf k} space differentiation of the spin excitations which has been studied later in great detail by ARPES experiments ({\bf section~\ref{Dynamic spin susceptibility and electronic correlations}}).  

\subsection{Carriers and Hyperfine couplings}
\label{Carriers and Hyperfine couplings}
Let us assume that the Cu holes responsible for the local moments yielding the AF state of the parent compounds and of the doped holes expected to be located on the oxygen orbitals are uncorrelated. In that case the macroscopic magnetic susceptibility should sum up the contributions of these two bands, while the $^{63}$Cu nuclear spins would probe the spin contribution on the copper sites. Similarly the $^{89}$Y and $^{17}$O nuclear spins would be more likely coupled to the oxygen holes. The determination of the hyperfine fields ({\bf NMREPS}) 
which couple the nuclear spins with the susceptibility has been essential in the understanding of the electronic structure. The anisotropies of orbital contributions to the $^{63}$Cu NMR shifts and of the $^{63}$Cu spin hyperfine couplings permitted to establish that the Cu holes are located in the Cu 3d$_{x^{2}-y^{2}}$ orbitals (Takigawa {\it et al} 1990). The evidence for a {\it negative} hyperfine coupling of $^{89}$Y with the spin susceptibility allowed to demonstrate that $^{89}$Y also probes the susceptibility localised on the Cu 3d$_{x^{2}-y^{2}}$orbitals through a transferred hyperfine coupling via O2p$_{\sigma }$ orbitals (Alloul {\it et al}, 1988), which was found identical for the insulating and doped compounds. This suggested that the spin susceptibility resides on a {\it single spin fluid} (Mila and Rice 1989), involving Cu 3d$_{x^{2}-y^{2}}$- O2p$_{\sigma }$ hybridized
orbitals, so that the two type of holes are correlated and not independent as would be suggested by {\bf Fig.~\ref{Fig4}(d)}. This is fully confirmed below by the analysis of the $T$ variations of the NMR shifts.
\subsection{Evidence for a pseudogap from NMR shift data}
\label{Evidence for a pseudogap from NMR shift data}
\begin{figure}
\centering
\includegraphics[height=8cm,width=10cm]{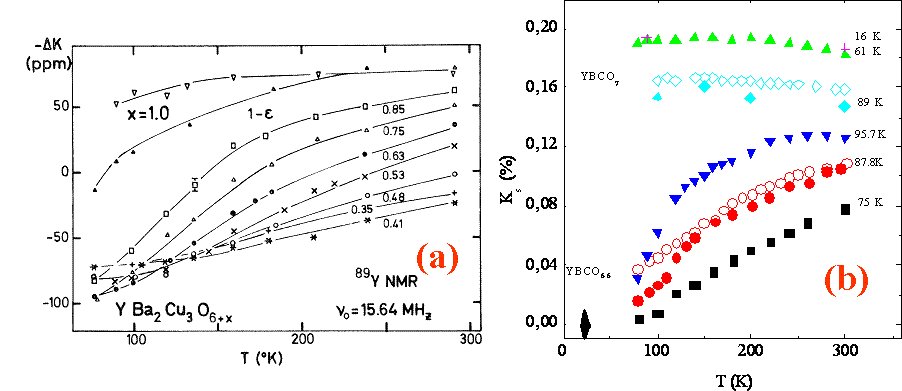} 
\caption{{\bf(a)} Temperature variation of the $^{89}$Y NMR shift $
-\Delta K_{s}$ for YBa$_{2}$Cu$_{3}$O$_{6+x}$ powder samples from optimal doping to a non-superconducting sample for $x=0.41$. The progressive increase of the pseudogap magnitude is apparent (from Alloul {\it et al}, 1989).{\bf (b)} The comparison of the $^{17}$O  NMR shift data in YBCO  and Hg1201 permits to demonstrate that the pseudogap temperature $T^{\ast }$ are identical for these two compounds (from Bobroff {\it et al} 1997).}
\label{Fig5}
\end{figure}

The optimally doped highest $T_{c}$ compounds exhibited a rather regular $T$ independent susceptibility together with a strange linear $T$ variation of resistivity above $T_{c}$. The possibility to control the hole doping in the YBa$_{2}$Cu$_{3}$O$_{6+x}$ cuprate by decreasing the oxygen content which is inserted in the intermediate planes between the CuO$_{2}$ planes permitted controlled NMR experiments in the underdoped regime for which $T_{c}$ drops with decreasing hole doping. Those experiments revealed a quite distinct behavior of the NMR shifts with a dramatic drop of the spin component $K_{s}$ that is of the spin susceptibility with decreasing $T$. Such an observation done initially by $^{89}$Y NMR measurements (see {\bf Fig.~\ref{Fig5}(a)}) remarkably revealed that for a composition with $T_{c}=60 $K, the spin susceptibility drops of more than a factor three between room $T$ and $T_{c}$ (Alloul {\it et al} 1989). As the spin susceptibility remains still sizable at $T_{c}$, this appeared as the signature of the opening of an imperfect gap which was qualified as a {\bf pseudogap} already in 1989. This is remarkable inasmuch as it was not experimentally possible to detect any further sharp decrease of the spin susceptibility below $T_{c}$. The other aspect which was revealed by these experiments is that the onset
temperature $T^{\ast }$ of the drop in $K_{s}$ increases with decreasing
doping. This had led to the indication that the pseudogap magnitude increases with decreasing doping that is with decreasing $T_{c}$.
Most other experiments measuring uniform macroscopic responses, such as
specific heat, planar resistivity $\rho _{ab}$, do detect an onset
at similar temperatures as that of $T^{\ast },$ (Timusk and Statt 2000), which is undoubtedly the highest temperature below which a detectable deviation with respect to the high $T$ Pauli like behavior occurs. Signatures for the pseudogap have been seen as well on optical absorption, photoemission (ARPES), or tunnel effect experiments.

\subsection{Universality of the pseudogap phase diagram}
\label{Universality of the pseudogap phase diagram}

\begin{figure}
\centering
\includegraphics[height=8cm,width=10cm]{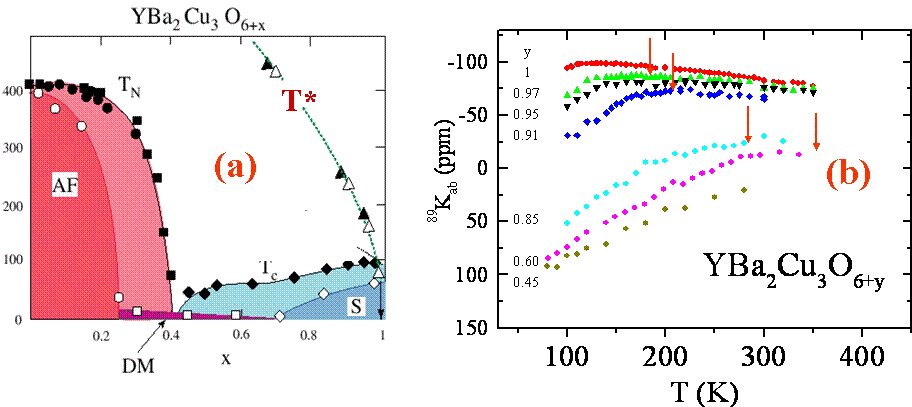} 
\caption{{\bf(a)} Cuprate phase diagram obtained in the YBa$_{2}$Cu$_{3}
$O$_{6+x}$ compound by changing the oxygen content $x$ of the Cu
intermediate planes. There the phase diagram obtained for Zn4%
substitution on the Cu sites demonstrates that $T_{c}$ is highly affected while $T^{\ast }$ values are insensitive to disorder.{\bf (b)} The determination of the $T^{\ast }$ values from the departure of the $^{89}$Y NMR shift from its high $T$ constant value is illustrated here by arrows. Figure composed from experimental results reported by Alloul {\it et al} (2009).}
\label{Fig6}

\end{figure}

Data taken for the spin components of the NMR shifts for $^{63}$Cu or $^{17} $O in YBa$_{2}$Cu$_{3}$O$_{6+x}$ have evidenced a perfect scaling of the $T$ variations with that of $^{89}$Y, which confirmed the idea of a single spin fluid contribution to the spin susceptibility. That was on line with the Zhang and Rice (1988) suggestion that oxygen holes just form singlets with Cu and only modify the Cu susceptibility, so that the Cu and O holes are highly correlated. This identical $T$ variation found by NMR on the various nuclear spin sites has given a universality to the pseudogap $T^{\ast }$ deduced by NMR. Comparison between $^{17}$O NMR shifts in the YBa$_{2}$Cu$_{3}$O$_{6+x}\ $ two layer compound and the single layer compound Hg$_{1}$Ba$_{2}$CuO$_{4}$ evidenced that $T^{\ast }$ is generic within the clean cuprate families (Bobroff {\it et al} 1997) (see {\bf Fig.~\ref{Fig5}(b)}). This has been confirmed by nearly all experimental determinations done by macroscopic measurements of $T^{\ast }\ $. This pseudogap $T^{\ast }\ $line introduced in the phase
diagram of YBCO is dispayed in {\bf Fig.~\ref{Fig6}} for pure samples but also when $T_{c}$ and $T_{N}$ have been decreased by $4\%$ Zn substitution on the Cu sites, as will be discussed in {\bf section~\ref{Exotic superconductivities}}.

\begin{figure}
\centering
\includegraphics[height=7.5cm,width=14cm]{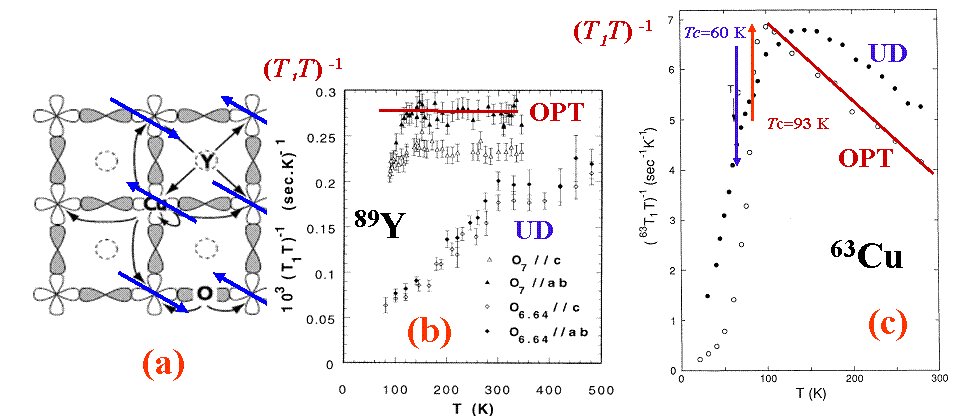}
\caption{{\bf(a)} Schematics of the Cu3d$_{x^{2}-y^{2}}$and O2p$ _{\sigma }$ orbitals in the CuO$_{2}$ plane showing why the AF fluctuations are filtered at the  $^{17}$O and $^{89}$Y nuclei. This explains why the $T$ variations of $(T_{1}T)^{-1}$ taken for oriented powder samples of YBCO$_{7}$ ($T_{c}=90$K) and YBCO$_{6.6}$ ($T_{c}=60$ K) are quite different for $^{63}$Cu and $^{89}$Y. {\bf(b)} $^{89}$Y data taken from (Alloul {\it et al} 1993).{\bf (c)} $^{63}$Cu data taken from (Takigawa {\it et al.} 1991). While the data for $T_{c}=90$K are $T$ independent for $^{89}$Y, as does the Knight shift, they increase at low $T$ for $^{63}$Cu. Similarly for the underdoped samples both $(T_{1}T)^{-1}$ and $^{89}K$ increase regularly up to $T^{\ast }$ $\sim350$ K, while those for $^{63}$Cu display a maximum at T $\sim150$ K,assigned to a spin gap.}
\label{Fig7}
\end{figure}

\subsection{Dynamic spin susceptibility and electronic correlations}
\label{Dynamic spin susceptibility and electronic correlations}
We shall discuss below the actual information on the AF correlations given by the measurements of the spin lattice $T_{1}$ and transverse $T_{2}$ nuclear spin relaxation in the cuprates. As shown in {\bf NMREPS}, the $T_{1}$ give determinations of $\chi''(q,\omega )$  while we shall show here that the transverse $T_{2}$ is related to ${\chi }^{\prime}(q,0)$.

The feature which had been clearly evidenced was that for $^{89}$Y nuclear spins $(T_{1}T)^{-1}$ and $^{89}$K have very similar $T$ variations (with $T$ and doping). This is illustrated in  {\bf Fig.~\ref{Fig7}(b)} on data taken on field-aligned YBCO samples realized for two compositions, O$_{7}$ for which $(T_{1}T)^{-1}$ is $T$ independent as is $^{89}K$, while for O$_{6.6}$ both quantities exhibit large $T$ increases. Similar results on the $^{17}$O NMR have been obtained, which established that the dynamic susceptibility viewed by these nuclei appeared quite correlated with the static susceptibility. It has been established that $T_{1}TK_{s}$ is nearly $T$ independent, which has been taken as an evidence for the presence of a Fermi-liquid like component in the magnetic response. 

However, as was seen by many authors, the $(T_{1}T)^{-1}$ of $^{63}
$Cu behaves quite differently (for references, see Walstedt, 2008). In the optimally doped compound $(T_{1}T)^{-1}
$ increases at low $T$ while it goes through a maximum at a temperature much lower than $T^{\ast }$ in the underdoped sample (see {\bf Fig.~\ref{Fig7}(c)}). This difference between $^{63}$Cu and $^{17}$O (or $^{89}$Y) NMR is understood as the two latter nuclear spins being coupled to two (or four) Cu moments do not detect AF fluctuations at the AF wave vector {Q}$_{AF}=(\pi ,\pi)$, as sketched in {\bf Fig.~\ref{Fig7}(a)}. In other words, the $^{63}$Cu data uniquely reveals the occurrence of a peaked response of $\chi''(q,\omega )$  at {Q}$_{AF}$. This has been confirmed directly by inelastic neutron scattering experiments taken on underdoped samples. The maximum in $(T_{1}T)^{-1}$ for $^{63}$Cu has been assigned to a spin gap which is quite distinct from the pseudogap $T^{\ast }$. It would increase much less rapidly than $T^{\ast }$ for decreasing doping. Both the pseudogap $T^{\ast }$ and the spin gap are detected only in underdoped samples, which suggests that they are connected.
 
Let us point out now that this strong magnetic response in cuprates
induces a contribution to the nuclear spin transverse $T_{2}$
relaxation, which has been found quite important on the Cu sites. In weakly correlated solids $T_{2}$, which is measured with spin echo
experiments (see NMR wikipedia) is usually fully determined by the direct dipole dipole interactions between nuclear spins . In
cuprates and more generally in correlated systems, a nuclear spin at $\mathbf{R_{i}}$ can be viewed as a moment which induces, through the {\bf q} dependent susceptibility ${\chi }^{\prime}(q,0)$, a polarization of the electronic spins which extends on the sites nearby $\mathbf{R_{i}}$. This polarization does in turn couple to the nuclear spins on these sites. This indirect (RKKY like) dipolar interaction between the nuclear spins induces a contribution to the spin echo decay. After summation of the interaction of a nuclear spin with all its neighbors, the spin echo is found to get a gaussian decay with a time constant $T_{2g}$ given by
\[
\label{T2g}\left(  \mathbf{1/T}_{\mathbf{2g}}\right)  ^{2}\mathbf{\ \propto
A}_{\mathbf{0}}^{4}\ \mathbf{\sum}_{\mathbf{q}}\mathbf{\left[  \mathbf{\chi
}^{\prime}(q,0)\right]  }^{2}\mathbf{-\ A}_{\mathbf{0}}^{4}\ \left[
\mathbf{\sum}_{q}\ \mathbf{\chi}^{\prime}(q,0)\right]  ^{2}.
\]
In the cuprates,  ${\chi }^{\prime }(q,0)$ is expected to be
peaked at $Q_{AF}$ and the width of the peak defines a correlation length ${\xi}$ for the AF response, which might be estimated from the $T_{2g}$ data. Even in the underdoped pseudogapped regime ${\xi}$ is found to increase steadily with decreasing $T$, as will be seen as well from impurity studies to be described in {\bf section~\ref{Magnetic properties induced by in-plane impurities in cuprates}}.

Coming back to the pseudogap, more recently ARPES or STM experiments have given evidence that a gap in the charge excitations only occurs for the antinodal directions $(0,\pi )$ in {\bf k} space. So the closed Fermi surface which occurs at high $T$  in underdoped cuprates loses weight in the antinodal directions when $T$ decreases, and the Fermi surface then reduces to Fermi arcs, which shrink with decreasing $T$ ( Kaminski {\it et al}, 2014). The experimental results on $(T_{1}T)^{-1}$
of $^{63}$Cu are certainly precursor indications of this {\bf k} space
differentiation which has been found by {\bf k} dependent spectroscopies. 

Phenomenological attempts have been done to describe the shape functions of the spin susceptibilities $\chi''(q,\omega )$ and $\ \chi ^{\prime }(q,0)$, in order to fit the NMR data (Millis {\it et al} 1990). Satisfactory qualitative descriptions could be achieved, with ${\xi}$ values of about two lattice constants at room $T$ in both the optimal and underdoped samples, with much larger low $ T$ increases of ${\xi}$ in the latter. However these approaches required to introduce by hand the Fermi liquid like metallic component and did not include explicitly the occurrence of the pseudogap. A complete theory of the physical
phenomena at play would require a model which generates altogether the
pseudogap, the AF correlation length and its $T$ variation. 

To conclude, the pseudogap is most probably intimately linked with the
correlated nature of these systems, and its actual physical origin is
intensely debated. One interpretation, proposed quite early on, is that it represents a precursor pairing state, the superconducting phase being only established at $T_{c}$ when the pairs achieve long range phase coherence (Emery and Kivelson 1995). Such an interpretation would imply that the SC gap increases with decreasing $T_{c}$. This is so far contradicted by direct or indirect determinations of the SC gap. Another class of interpretations could be the establishment of a hidden order disconnected from superconductivity such as a spin ordering, for instance a Resonant Valence Bond (RVB) state (Anderson,1987), a d-density wave (ddW), a charge segregation into stripe order or an ordering involving orbital currents. Such possibilities have been
recently underlined by experimental discoveries of such type of orders, which appear system dependent, and often occur at temperatures below $T^{\ast }$.These experiments are so novel that they have initiated vivid debates on the pseudogap, but did not permit so far to resolve the issues they raised. The pseudogap remains still today the central point debated on the cuprates and at the present writing the understanding
of the pseudogap state remains controversial. The author believes that
magnetic short range correlations explain the pseudogap crossover at $
T^{\ast }$ and the Fermi surface differentiation, while the orders detected at lower $T$ than $T^{\ast }$ are rather consequences of the pseudogap formation than direct manifestations of the pseudogap itself.\\

{\bf The analysis of the hyperfine couplings established that the magnetic response of cuprate High $T_{c}$ superconductors exhibit a single spin fluid behavior in which the Cu and O holes are hybridized. NMR shift experiments have allowed to evidence that a large progressive loss of spin susceptibility occurs for underdoped cuprates, that is for compounds on the left of the $T_{c}$ dome. This strange behavior has been found generic for the various cuprate families. It onsets at a temperature $T^{\ast } >> T_{c}$ and is attributed to a pseudogap in the density of states which corresponds to a transfer of spectral weight from low to high energies, assigned to strong electronic correlations.  The difference of $T$ dependence of $T_{1}$ for the nuclei in symmetric position in the Cu site lattice with respect to that of Cu nuclear spins confirms the existence of AF correlations between Cu electronic spins. Modifications in the Fermi surface topology have been confirmed by k dependent spectroscopy and various further reconstructions of the Fermi surface have been discovered to occur well below $T^{\ast }$.}

\section{1D metals and Luttinger Liquids}
\label{1D metals}
Lattice dimensionality gives specific effects, the case of 1D metals or spin chains being certainly one of the most original. Indeed 1D compounds have a very simple {\it Fermi Surface} which is restricted to two points at $k_{F}$ and -$k_{F}$. This delineates a singular behavior, as in such a condition low energy electron-hole excited states can only occur for $q\simeq0$ or $q\simeq2k_{F}$. For higher dimension electron hole pairs with a given energy  can be produced with any difference in $q$ values. The peculiarity in 1D is therefore that the excited quasiparticles can by no way be described as Fermi liquid quasiparticles and that, whatever the strength of the electronic interactions.

Another peculiarity of the 1D systems is that the Fermi surface always displays nesting properties near the Fermi level, which can be easily seen by linearising the dispersion relations near $\pm k_{F}$, so that electronic instabilities yielding CDW or SDW ordered states are easily favored. Furthermore one could notice that in 1D  the displacement of an electron cannot occur without a concomitant motion of all other electrons along the chain, so that any transport drives a collective excitation. The incidence of electronic interactions is therefore maximal in such systems. The peculiar electron hole pairs in 1D systems do still have a well defined width and disperion relation, and are therefore quasiparticles which constitute a Luttinger liquid (LL) quite distinct from a Fermi liquid. 

Let us point that the excitations at $q\simeq0$ have quite distinct significance than those at $q\simeq2k_{F}$. Those at $q=0$ determine the spin susceptibility and correspond to a limiting constant DOS at low $T$. They behave similarly as for a Fermi liquid and give for instance in NMR a Korringa $T$ linear contribution to the spin lattice rate $(T_{1})^{-1}$. But the excitations at $q\simeq2k_{F}$ are quite characteristic of the Luttinger liquid and involve both charge and spin excitations. The physical observables in this Luttinger liquid are driven  by correlation functions which display specific temperature exponents which are directly related to the strength of the electronic correlations. Those can be determined theoretically (Giamarchi 2004), and for instance the nuclear spin lattice relaxation rate is not governed at low $T$ by the Korringa law $1/T_{1}\propto T$ but corresponds to $1/T_{1}\propto T^{\alpha }$, the
value of the exponent $\alpha $ being connected then with the strength of the electron electron interactions, that is with $U$. One expects to see a change in the nuclear relaxation exponents from a Korringa law to an $\alpha$ exponent at low $T$. By combining NMR and transport experiments one can get the exponent controlling both spin and charge correlations (Behnia {\it et al},1995).
However in 1D systems charge density wave (CDW) or  spin density wave (SDW) ordered states compete with the Luttinger liquid state, so that the actual situation is such that in many practical cases the 1D Luttinger behaviour is not maintained at the lowest temperatures. Among the various competing ground states one finds as well Peierls dimerization (see {\bf NMREPS}), or a crossover towards a 2D behaviour induced by the transverse interactions between chains. The description of such complicated situations would require lengthy developments on the physics of 1D systems which would go well beyond the scope of the present article.
\begin{figure}
\centering
\includegraphics[height=5cm,width=8cm]{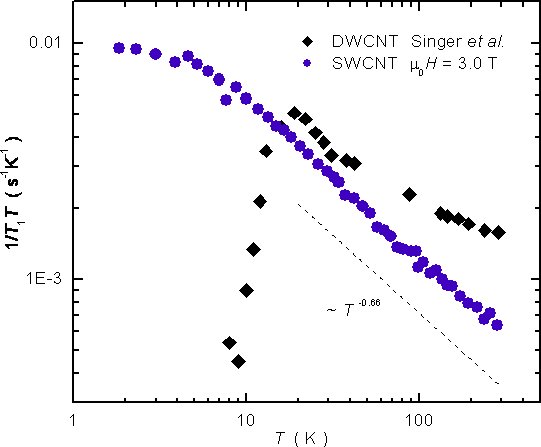}
\caption{$(T_{1}T)^{-1}$ data for the $^{13}$C nuclear spins
display a $T^{-0.66}$ behavior in SWCNT on a wide $T$ range from 6 to
300 K, which would correspond to a Luttinger exponent of $\alpha =0.33.$
(Ihara {\it et al}, 2010b). For DWCNT, in which $^{13}$C enrichment of the inner tubes had been achieved, a low $T$ gap in the excitations of the inner tubes is detected (from Singer {\it et al} 2005).}
\label{Fig8}
\end{figure}

Hints for such a LL state could be seen at high $T$ in the organic TMTSF compounds (Wzietek {\it et al}, 1993), which are quasi 1D materials. However in these compounds the LL state could not be followed toward low $T$ as it is obscured by the existence of 2D competing magnetic states which are gapped at low $T$. As will be seen in {\bf section~\ref{Metallic magnetic states and Mott insulator to metal transition}} these compounds display phase diagrams with a proximity between magnetic and SC phases as for the cuprates. However in such quasi 1D compounds the number of carriers is difficult to modify, and the electronic properties are rather spanned by changes of transfer integrals induced by an applied pressure. A LL behaviour has been better evidenced by NMR from $1/T_{1}$ measurements on $^{13}$C in single wall carbon nanotubes (SWCNT). Such nanotubes were expected to display one dimensional behaviour (Giamarchi 2004). The NMR data shown in {\bf Fig.~\ref{Fig8}} does indeed evidence a power law exponent on a large $T$ range. Conversely $^{13}$C NMR data taken on the inner tubes of double wall tubes (DWCNT) rather exhibit a metallic state at high $T$ with a gap opening at low temperature, which is not really understood so far. It could be speculatively assigned to a loss of 1D behaviour associated with nanotube curvature effects. 
For completeness let us mention a spin system of reduced dimensionality (C$_{5}$H$_{12}$N)$_{2}$CuBr$_{4}$ which can be considered as formed of nearly decoupled chains of singlet dimers. In those 1D chains the triplet excitations of the dimers have been shown to form a Luttinger Liquid (Giamarchi 2004).\\

{\bf In 1D systems the peculiarity of the Fermi surface which is restricted to two points, and of the atomic structure, yields a breakdown of Fermi liquid theories. The correlated electron physics is described in that case by a Luttinger liquid theory with correlation functions which display singular exponents for the various physical quantities. Those are observed for instance in $1/T_{1}$ experiments taken on single wall carbon nanotubes.}

\section{Impurities and local responses in correlated electron systems}
\label{Impurities and local responses in correlated electron systems}
An impurity is a local screened Coulomb potential, which ideally is a
uniform perturbation in {\bf q} space, inducing a response which is
inhomogeneous in real space but which reflects the response to all {\bf q}values. So quite generally an impurity potential is a fundamental tool to probe the specific response of a pure system. For instance, the RKKY
oscillations induced by a local moment impurity in a classical metal are
due to the singularity associated with the truncation of the response at $\left\vert \mathbf{q}\right\vert =k_{F}$. In correlated electron systems, if some singularity occurs at a specific '''q'''  value for a pure system, this singular response will dominate the modifications introduced by the impurity. For instance, in magnetic materials for which AF correlations can be important at a wave vector $\mathbf{q}=\mathbf{q}_{AF}$, a staggered magnetic response at this wave vector is expected (as shown in {\bf Fig.~\ref{Fig9}}). So quite generally an impurity potential is a fundamental tool to probe the specific response of a pure system, which is all the more interesting when some  physical properties might be hardly measurable directly in the pure system, or when some of the hidden physical properties can be revealed by the impurity potential. NMR experiments ideally permit to map out the spatial changes occurring around extrinsic defects. Those cannot be accessed through macroscopic non-local techniques. Here we shall evidence that in spin chains the staggered magnetism induced by
defects in spin chains allows to determine by NMR the correlation functions of the {\it pure} state ({\bf section~\ref{Haldane chains}}). Similarly in the cuprates Zn or Li non magnetic atoms substituted on the Cu sites were shown to induce an extended paramagnetic state in their vicinity ({\bf section~\ref{Magnetic properties induced by in-plane impurities in cuprates}}). Such studies have been important to qualify the incidence of disorder in the various doping ranges of the cuprate phase diagram ({\bf section~\ref{Influence of disorder on the cuprate phase diagram}}) (see Alloul {\it et al.} 2009).

\subsection{Haldane chains: NMR of sites near impurities}
\label{Haldane chains}
\begin{figure}
\centering
\includegraphics[height=8cm,width=13cm]{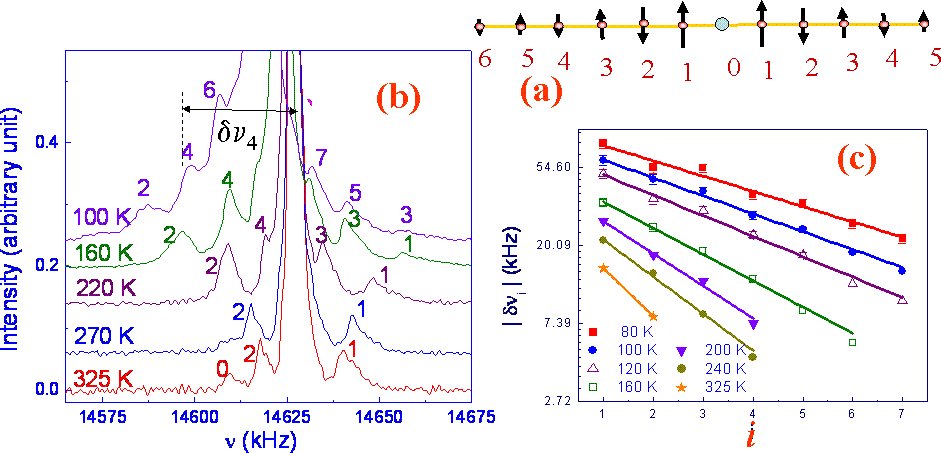}
\caption{{\bf(a)} Sketch of a chain with an impurity or a missing site. A paramagnetic perturbation is induced by the defect.{\bf(b)}$^{89}$Y NMR spectra taken on the Y$_{2}$BaNiO$_{5}$ Haldane chain compound (see text). They display a central line signal and satellite lines on both sides of the central line (from Das {\it et al.} 2004). {\bf (c)} The analysis of the $T$ dependence of the spectra reveals the $T$ variation of the impurity induced correlation length.}
\label{Fig9}
\end{figure}

In actual chain materials the chains are never of infinite length so a chain end is a defect which induces a magnetic response on the chain. A
quantitative study of these effects can be done by introducing a controlled concentration of impurities which disrupt the chains ({\bf Fig~\ref{Fig9}(a)}). For instance Zn, Mg with $(S=0)$ and Cu with $(S=1/2)$ have been used as substituents at the Ni site of the Y$_{2}$BaNiO$_{5}$ compound in which the $S=1$ spins at the Ni sites form a chain (this is typical case of an integer spins chain, named as Haldane chain). The $^{89}$Y NMR spectrum consists of a central peak and several less intense satellite peaks which permit to probe the staggered magnetization (Teloldi {\it et al}, 1999). In the NMR spectra of {\bf Fig.~\ref{Fig9}(b)} the central peak corresponds to chain sites
far from the defects and its NMR shift measures the uniform susceptibility, which displays a Haldane gap $\sim 100$  K which corresponds to an antiferromagnetic (AF) coupling $J\approx 260$ K between the nearest neighbor Ni spins. The satellite lines are evenly distributed on the two sides of the central peak. The magnitude of their shifts $\delta \nu _{i}(T)|$ with respect to the main line varies exponentially as a function of distance to the impurity at any given $T$ ({\bf Fig~\ref{Fig9}(c)}). From fits with 
\[
|\label{corr length}\delta \nu _{i}\,(T)|=|\delta \,\nu _{1}(T)|e^{-\frac{(i-1)}{\xi _{imp}(T)}}
\]

one deduces the extension $\xi _{imp}(T)$ of the induced staggered
magnetization. The values of $\xi _{imp}(T)$ extracted from these
measurements are found almost identical to the pure correlation length
computed by Monte Carlo simulations for an infinite $S=1$ chain with no
defects. The characteristic length of the staggered magnetization near defects therefore reveals the intrinsic $\xi $ of the infinite $S=1$ chain, in the explored $T$ range (Teloldi {\it et al.}, 1999). Furthermore the $T$ dependence of the induced paramagnetic moment has been found to display a Curie free spin behavior.

\subsection{Magnetic properties induced by in-plane impurities in cuprates}
\label{Magnetic properties induced by in-plane impurities in cuprates}
\begin{figure}
\centering
\includegraphics[height=8cm,width=10cm]{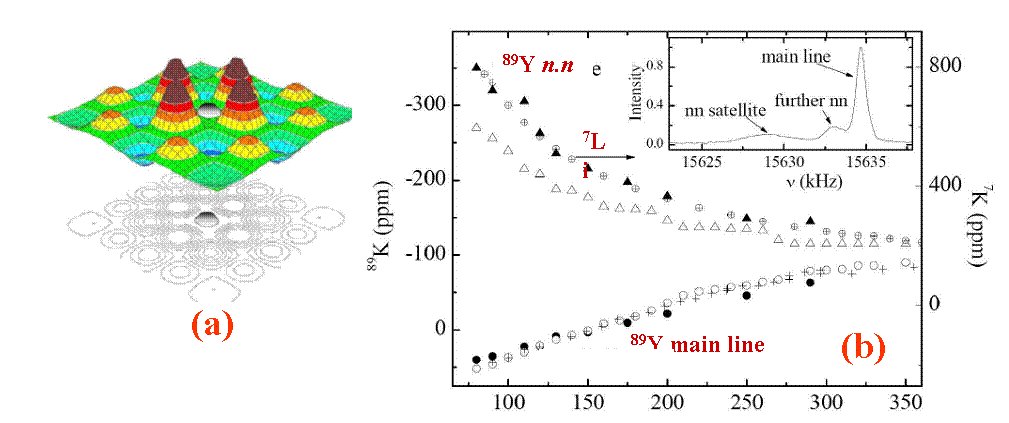}
\caption{{\bf(a)} Non magnetic impurities such as Zn or Li substituted on the Cu site in the CuO$_{2}$ plane in cuprates induce a 2D staggered magnetic response.{\bf(b)} {\bf inset}: The $^{89}$Y NMR spectrum exhibits a central line and satellite lines as in spin chains. {\bf Main}: the $T$ dependence of the satellite or $^{7}$Li NMR shift permits to monitor that of the induced paramagnetism. (Figures from Alloul {\it et al.} 2009)}

\label{Fig10}
\end{figure}

As shown above in {\bf section ~\ref{Magnetic imputrities & Kundo effect}} the underdoped regime of cuprates, for which a pseudogap is detected, is the interesting range where the system is a metal with magnetic correlations, for which the use of impurities to probe the physical properties was expected to be the most fruitful. The first basic qualitative information were obtained by an approach, started in
the early 1990's, using Zn and Ni impurities substituted for Cu in the YBaCuO$_{6+x}$ system specifically for $x\simeq 0.6$ for which the
pseudogap occurs at $T^{\ast }>>T_{c}$. The question which arose was whether a nonmagnetic site induces a free paramagnetic moment in a {\it metallic} correlated system, as was seen in the case of undoped spin chains and ladders. An indirect but unambiguous evidence that Zn induces a paramagnetic moment in an underdoped cuprate was obtained by monitoring the $^{89}$Y NMR linewidth in YBCO$_{6.6}$:Zn$_{y}$ (Alloul {\it et al.} 1991). The significant low $T$ increase of the linewidth that was detected revealed the increase of the static staggered spin polarization around the impurity. This was clearly confirmed later by resolving in dilute samples the satellite NMR signals of the $^{89}$Y near neighbor ({\it n.n.}) sites of the substituted Zn (Mahajan {\it et al.} 1994) (see {\bf Fig.~\ref{Fig10}(b)}). This provided the first local detection of the field induced paramagnetism near the Zn, well before the equivalent information could be monitored in the case of spin chains. These data implied that the spin polarization of the Cu ''n.n.'' to Zn is already at 100 K more than ten times larger than that of the pure host, so this was not a mere minor modification of the host density of states, but a strong effect of the electronic correlations of the system, similar to that found for spin chains. 
Quite analogous $^{89}$Y {\it n.n.} NMR data were obtained later (see {\bf Fig.~\ref{Fig10}(b)} for
non magnetic Li impurities, which provided the possibility to use in the same sample the $^{7}$Li, $^{89}$Y, $^{17}$O and $^{63}$Cu nuclear spin probes. The $^{7}$Li NMR permitted accurate measurements of $\chi (T)$ of the four Cu {\it n.n.} of the Li non magnetic impurity. In the underdoped samples, this variation was found to display a
Curie variation at low doping, which confirmed the observation made from the $^{89}$Y NMR that the impurity induced state behaves as a nearly free paramagnetic moment (Bobroff {\it et al.} 1999). For increasing doping the Curie law is found to transform into a Curie Weiss law with a Weiss temperature $\ \Theta $ which increases abruptly with doping. One could conclude that the low $T$ reduction of susceptibility in the optimally doped case is due to the onset of the energy scale $k_{B}\ \Theta $ in analogy with the Kondo reduction of local moments in classical metallic systems. The data taken on the other nuclei has enabled the quantitative
determination of the spatial structure of the induced polarization, that is its magnitude and $\xi _{imp}$ $(T)$ which increases
significantly at low $T$. Although $\xi _{imp}$ has been found of similar magnitude at room temperature for optimal doping it dsplays much weaker variations at low $T$ than in the underdoped case. The energy scale $\Theta $ may control the $T$ variations of both quantities, however. Since overdoping corresponds to an increase of $\Theta $ well beyond the value found for optimal doping, such a scheme
would allow a smooth crossover towards the Fermi liquid limit for large
overdoping. Data on the spin lattice relaxation of $^{7}$Li allowed to study the $T$ dependence of the relaxation rate $1/\tau _{s}$ of the local moment staggered electronic spin induced in the CuO$_{2}$ plane. The behavior of $1/\tau _{s}$ has been found quite similar to that found for $1/\chi _{c}$ and provided direct evidence that
$\chi _{c}$ and $\;\tau _{s}^{-1}$ are governed by the same energy scale. In other words, the applicability of a Curie-Weiss law for $\chi _{c}$, with a Weiss temperature $\Theta \;$is associated with a limiting low $T$ behavior $\tau _{s}^{-1}\;=k_{B}\Theta /h\;$ for the local moment fluctuation rate. This experimental evidence is strikingly reminiscent of that encountered for a local moment in a noble metal host ({\bf section ~\ref{Magnetic imputrities & Kundo effect}}), the Kondo energy being then the width of the resonant state which governs at low $T$ the fluctuations in the Kondo state (Alloul {\it et al} 2009).

\subsection{Influence of disorder on the cuprate phase diagram}
\label{Influence of disorder on the cuprate phase diagram}
\begin{figure}
\centering
\includegraphics[height=8cm,width=10cm]{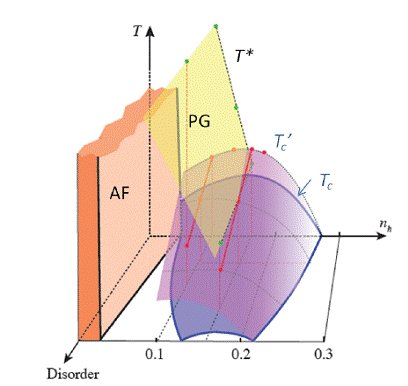}
\caption{Extension of the ($n_{h},T$) cuprate phase diagram to 3D, using disorder as a third parameter. The green and red points are data taken for YBCO$_{6+x}$ with varying disorder. There the pseudogap (yellow) is not modified by disorder, while the AF range (orange) is only slightly reduced. The superconducting dome (blue) shrinks markedly and gives room to a magnetic disordered phase which separates the AF and SC phase (see {\bf Fig.~\ref{Fig6}(a)} for the case of Zn substitution.  Finally the extension $T'_{c}$ of the SC fluctuations (light violet) follows the $T_{c}$ dome, but its extension with respect to $T_{c}$ increases for strong disorder. (from Rullier-Albenque {\it et al.} 2011).}

\label{Fig11}
\end{figure}
 
Studies of impurities in correlated electron systems are also important to help us understand the influence of the intrinsic disorder that is present in nominally pure systems. In cuprate and  heavy fermion materials, one can easily change the carrier concentration and thus explore a wide range of the phase diagram on the same compound. However the very doping process introduces disorder. The importance of this phenomenon has only been recently appreciated by much of the scientific community, as local techniques such as STM have imaged impurity states directly and brought to the fore the study of disorder as a key issue in correlated electron systems. The efforts which have been made so far with NMR to understand the influence of controlled disorder and impurities are therefore very good reference points to try to
understand the incidence of intrinsic disorder. 

We wish to stress that in the intermediate doping range substitution impurities interfere with the intrinsic disorder of the {\it pure}
systems. So one has to consider that non magnetic impurities increase the existing disorder in these phases. For instance one should have noticed in {\bf Fig.~\ref{Fig6}(a)} that Zn substitution in
YBaCuO extends significantly the disordered magnetism range in this material, so that the phase diagram of YBaCuO:Zn$_{0.04}$ resembles that observed in nominally pure La$_{2-x}$Sr$_{x}$CuO$_{4}$, or in Y$_{1-y}$Ca$_{y}$BaCuO$_{6}$. This suggests that extrinsic or intrinsic disorder may have similar influences on the properties of the cuprates
at least in the low hole doping range.

The most important point which has been intentionally investigated in the early experiments by NMR was the influence of the substituted impurities on the temperature $T^{\ast }$ at which the pseudogap opens. As could be seen on $^{89}$Y NMR data, the shifts of nuclear spins located far from the Zn impurities are not affected,even when the impurity content is large enough to completely suppress superconductivity (Alloul {\it et al} 1991). The absence of experimental modification of the pseudogap line in {\bf Fig.~\ref{Fig6}(a)} has been confirmed also for Li, Ni impurities, irradiation defects
and any kind of disorder, with most probes which can determine the normal state pseudogap far from the defects. Overall the experiments give consistent evidence that strong modifications are only induced locally around the defects. 

Therefore, we have been led to conclude that one should not consider that a universal phase diagram occurs in the ($n_{h},T$) plane for the cuprates. Intrinsic or extrinsic disorder induces changes on the actual physical properties, which are reflected in the doping dependence of the phase diagram. We have suggested that this can be taken into account in an extension to 3D of the phase diagram in which the third axis represents the disorder as an extra parameter ({\bf Fig.~\ref{Fig11}}). One can see there that the extension of the magnetic disordered phase
between the AF state and the SC dome increases with disorder, as evidenced in {\bf Fig.~\ref{Fig6}(a)}. In this sketch of a phase diagram the pseudogap line remains insensitive to disorder. The onset of SC fluctuations $T'_{c}$, as determined from the SC fluctuations contribution to the conductivity is also represented there. It is seen that $T'_{c}$ is only slightly above the SC dome and is less modified than $T_{c}$ by an increasing disorder. In such a representation one can see that the variation of optimal $T_{c}$ and the various phase diagrams encountered in specific cuprate families can be tentatively associated with distinct amounts of disorder. They would be generated by vertical cuts taken for fixed disorder.\\

{\bf The spatial dependence of the magnetic properties of correlated electron systems can be revealed by NMR studies of the response induced by impurities in the correlated material. This has been applied for instance in the case of spin chains or in the cuprate superconductors. The NMR of near neighbours of the impurities permits to determine the magnetic correlation length characteristic of the pure material and its temperature dependence. This also leads to consider that impurities and defects do influence certain features of the phase diagram of correlated electron systems. In the cuprates this has established that the pseudogap is independent of disorder while the transition from AF to SC is quite dependent of intrinsic or extrinsic disorder.}

\section{Exotic superconductivities}
\label{Exotic superconductivities}
The importance of the cuprates in the physics of correlated systems has
resulted from the discovery that when the AF is suppressed by hole doping, the doped metallic state which results has a SC ground state and displays strange metallic and magnetic properties. The most surprising feature has been the fact that the superconductivity discovered in these materials has the highest critical temperatures $T_{c}$ found so far in any superconducting material, and exceeds any $T_{c}$ which could be expected within the BCS approach known to apply in classical metallic states. An important observation in the cuprates has been the fact that the phase diagram with increasing hole doping displays a dome shaped SC regime, that is SC disappears for dopings beyond about 0.3. These non expected features have immediately led to the idea that SC in the cuprates has an exotic origin linked with electron electron interactions rather than the classical electron phonon driven superconductivity which prevails in classical metals.

\begin{figure}[h]
\centering
\includegraphics[height=6.5cm,width=12cm]{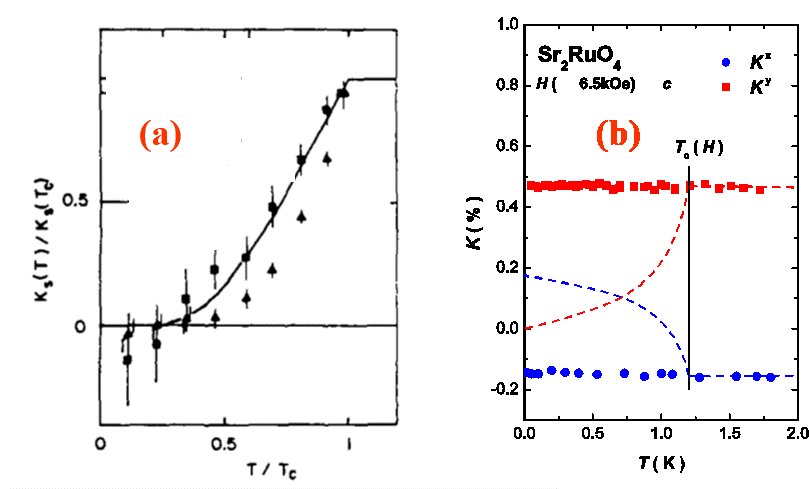}
\caption{$^{17}$O NMR shift data taken below $T_{c}$ in the two planar directions {\bf (a)} in YBaCuO${_7}$ it drops below $T_{c}$ and vanishes for $T<<T_{c}$ (Takigawa {\it et al.} 1989). {\bf(b)} in Sr$_{2}$RuO$_{4}$ the $^{17}$O NMR shifts remain constant through $T_{c}$, which supports spin triplet Superconductivity. Here the reported lines would correspond to expectations for a singlet SC case (from Ishida {\it et al.} 1998).}
\label{Fig12}
\end{figure}

Obviously, the establishment of any SC state yields profound transformations of the electronic properties which are reflected in the NMR response. In BCS Superconductors the formation of singlet Cooper pairs is directly seen as a loss of the normal state spin susceptibility that is a drop of the NMR shift ({\bf NMREPS}). NMR studies appeared then quite important in the early days after the discovery of HTSC. One indeed was interested to see whether BCS like observations would be done. For HTSC samples with high $T_{c}$ around the optimal doping, the NMR data appear quite similar to those obtained in standard BCS materials inasmuch as the NMR shift of most nuclear species in the material $^{63}$Cu, $^{17}$O, $^{89}$Y  were found $T$ independent down to $T_{c}$, and dropped abruptly at $T_{c}$ in accord with spin singlet superconductivity {\bf Fig.~\ref{Fig12}(a)} (Takigawa {\it et al.} 1989).  In many cases for which SC is probably also more exotic than for phonon mediated SC the pairing state remains singlet, which has been confirmed by  similar NMR shift studies.

In some exotic SC states the pair wave function can be in a spin triplet state, a situation which has been found first for superfluidity of $^{3}$He which are fermions which bind to form spin triplet Cooper pairs (Leggett 1975).  As a spin triplet can be in three distinct states either $|\uparrow \uparrow \rangle,|\downarrow \downarrow \rangle,|\uparrow \downarrow \rangle + |\downarrow \uparrow \rangle$ or quantum superpositions of these components, the magnetic response to an applied field depends of the actual state of the bound pairs. This explains why $^3$He has two different triplet superfluid phases, that called A with equal spin states formed by $|\uparrow \uparrow \rangle$ and $|\downarrow \downarrow \rangle$ pairs displays no change of the nuclear spin susceptibility though the superfluid transition, while phase B is an equal superposition of the three states which leads to a marked decrease of the spin susceptibility, which however does not vanish totally at $T=0$.
 
In correlated electron systems one similarly expects that with spin triplet SC the behavior of the NMR shift below $T_{c}$ should permit to establish the spin triplet pairing and to determine the superposition of spin states. A system which has been thoroughly studied is Sr$_{2}$RuO$_{4}$, in which the RuO$_{2}$ form a square lattice which is similar to that of the cuprate La$_{2}$CuO$_{4}$. Indeed it has been shown in that compound that both $^{17}$O and $^{99}$Ru  NMR shifts exhibit absolutely no change though $T_{c}$, {\bf Fig.~\ref{Fig12}(b)} which is a strong case for spin triplet SC with equal spin states (Ishida {\it et al.} 1998). Once the spin properties of the pairs has been established, their orbital state has a symmetry which is imposed by the total antisymmetry of the wave function. So that an antisymmetric spin singlet state implies an even orbital state, that is an $s$ or $d$-wave symmetry of the wave function. Similarly for a  symmetric triplet state, the orbital wave function should be antisymmetric that is $p$-wave or higher order. In most of these exotic pairing states the SC gap is not uniform over the Fermi surface as is the case for most phonon mediated cases. In these exotic superconductors the gap depends of the wave vector ({\bf k, -k}) of the pairs and might exhibit gap nodes for some {\bf k} values or some wave vector directions. For instance for a 2D system, if the gap has a $d$-wave wave order parameter symmetry, the gap changes sign and therefore vanishes along two axes of the unit cell. This implies that the gapless states will be filled much faster with increasing  $T$ than in a pure $s$ wave BCS superconductor.
 
\begin{figure}
\centering
\includegraphics[height=8cm,width=10cm]{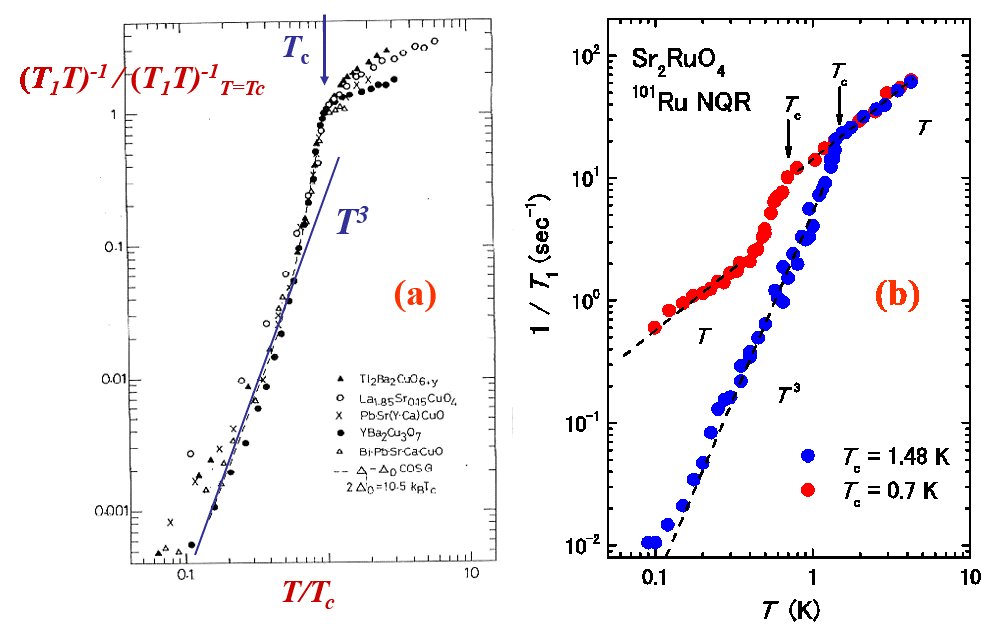}
 \caption{{\bf(a)} In YBaCuO7 the $^{63}$Cu $1/T_{1}$ data has a $T^{3}$ dependence, which agrees with {\bf d} wave SC (Asayama {\it et al.} 1991). {\bf(b)}In Sr$_{2}$RuO$_{4}$ the sample dependence of the Ru NQR $1/T_{1}$ data is illustrated. The use of a clean sample permits to evidence a $T^{3}$ variation which establishes the existence of nodes in the gap function (Ishida {\it et al.} 2000).}
\label{Fig13}
\end{figure}

In such spin singlet states the functional form of the increase of the spin susceptibilty (that is of the NMR shift) with increasing temperature from $T=0$ permits in principle to determine whether nodes occur in the gap function. Experimentally this is somewhat difficult to establish from NMR shift measurements which have limited accuracy at low $T$ due to the inhomogeneous field penetration in the vortex lattice ({\bf RMNEPS}). This is however much more accessible from $1/T_{1}$ data which display then a power law increase with an exponent which depends of the wave vector dependence of the gap. A $T^{3}$ variation of $1/T_{1}$ has been best evidenced by zero field NQR experiments in cuprates, which is in accord with the $d$ symmetry of the SC order parameter, {\bf Fig.~\ref{Fig13}(a)} (Asayama {\it et al.} 1991).
Though these NMR data were rather conclusive, this $d$ wave symmetry has such an implication for the understanding of the pairing mechanism that it has only been fully accepted within the community when ARPES and phase sensitive tunneling experiments established it independently.
One has however to recall that low $T$ NMR measurements can be contaminated by extra contributions of impurities to the relaxation. So to conclude about the symmetry of the SC order parameter in a given compound a great care has to be taken to avoid the presence of impurities in the actual materials. This is for instance illustrated in {\bf Fig.~\ref{Fig13}(b)} for the spin triplet SC of Sr$_{2}$RuO$_{4}$, for which a $T^{3}$ variation of $1/T_{1}$ has also been found 
once clean samples could be produced (Ishida {\it et al.} 2000).
Here again this result points out the existence of lines of gap nodes, though their spatial location on the Fermi surface is not yet clarified. The spatial structure of the $p$ wave state symmetry which governs SC in this compound is therefore not yet fully characterized. A $p$ wave spin triplet state has been proposed as well from NMR experiments in some Heavy Fermion compounds such as UPt$_{3}$ or in low D organic conductors, although more experimental confirmations would be required to fully establish the validity of these proposals.\\
 
{\bf In this section, we have indicated how NMR shift experiments below $T_{c}$ permit to establish the actual spin state of the paired particles in the superconducting state. NMR $1/T_{1}$ data taken at low $T$ can give some hints on the orbital state of the pairs and the actual order parameter symmetry. They can permit to establish the existence of nodes in the gap function. NMR experiments are therefore quite important for the characterization of non conventional superconductivity.}

\section{Charge differentiation and orbital order}
\label{Charge differentiation and orbital order}
In many transition metal oxides one usually expects a homogeneous behavior, that is the same electronic properties on all 3$d$ (or 4$d$) sites in the unit cell. However in many cases structural changes of the unit cell might happen at given temperatures and modify the electronic structure, which in turn differentiates the sites. We have shown in {\bf section~\ref{1D metals}} that this is a quite specific situation encountered in some 1D compounds for which a Peierls transition might lead a dimerization of the chains, which opens an insulating gap at low $T$. In 2D or 3D systems one might equivalently get changes in the electronic structure associated with Fermi surface reconstructions at low $T$ and stabilization of charge density waves (CDW).
Such differentiation of the sites can be probed ideally by NMR/NQR
techniques. As recalled in {\bf NMREPS} that can be done using nuclear
spins with spin $I>1/2$ for which quadrupolar effects are sensitive to the distribution of charges around the nuclei and allow the determination of Electric Field Gradients (EFG). Those involve a contribution from the distant ionic charges that can be calculated in a point charge model. This gives approximate estimates of the EFG. They turn out to be more reliable in insulators rather than in metals. But
the on site orbitals on the transition element also induce a sizable
contribution if all the $d$ levels are not equally populated. The
quadrupole frequency $\nu _{Q}$ contains these two contributions which we may write
\[
\label{quadrupole}\nu _{Q}=\nu _{Q\ }^{latt}+\ \nu _{Q\ \ }^{el},
\]
where $\nu _{Q\ }^{latt} $results from the distant ionic charges while
the local charge contribution $\nu _{Q\ \ }^{el}$ is due to the partly
filled $d$ orbitals, which could be either localized or involve the
carriers. One can therefore differentiate by NMR spectra the
sites which are in different atomic positions in the unit cell, or the sites with atomic orbitals involved in different bands of the electronic
structure. In those cases the charge occupancy differ on given sites,
which corresponds to disproportionation within a chemistry terminology.

\begin{figure}[h]
\centering
\includegraphics[height=7cm,width=9cm]{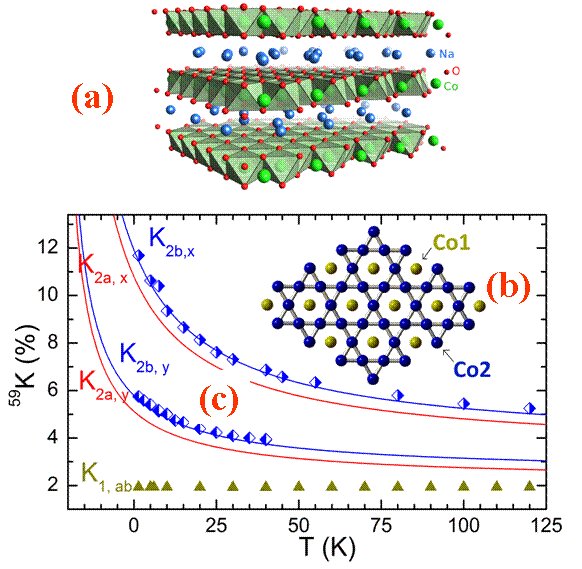}
\caption {{\bf(a)} Layered structure of Na$_{x}$CoO$_{2}$.{\bf(b)} 
the NMR data  for $x=2/3$ permits to evidence the disproportionation of two types of Co sites, the Co1 (yellow) and Co2 (blue).{\bf(c)} The $T$
dependence of the $^{59}$Co NMR shift is fully assigned to the spin
susceptibility on the Co2 sites which form a Kagome lattice on which
the holes are delocalized, while the Co1 complementary sites in the structure are non magnetic Co$^{3+}$ (Mukhamedshin and Alloul, 2011).} 
\label{Fig14}
\end{figure}

We shall exemplify here these possibilities as revealed by extensive NMR
experiments which have been dedicated to the study of the phase diagram of sodium cobaltates Na$_{x}$CoO$_{2}$. Those are layered oxide materials somewhat similar to the cuprates inasmuch as the charge doping of the CoO$_{2}$ layers is controlled by the Na content on a large range of concentrations. The significant difference of cobaltates with the cuprates is that the Co of the CoO$_{2}$ plane are ordered on a triangular lattice and not on a square lattice (see {\bf Fig.~\ref{Fig14}(a)}). 
In this configuration, the large crystal field on the Co site favors low spin states of Co ions. Their orbital degeneracy influences significantly the electronic properties. A rich variety of physical properties ranging from ordered magnetic states, large thermoelectric effect, high Curie-Weiss magnetism and metal insulator transition, superconductivity etc, have been observed in these compounds. Contrary to the case of most cuprates for which dopant induced disorder is quite influential, the doping achieved in cobaltate samples is associated with the insertion of well ordered 2D Na structures. This has been illustrated in {\bf NMREPS} by $^{23}$Na NMR/NQR experiments. Those have given evidence that for $x\geq0.5$ the Na atoms display ordrered states for all single phase samples synthesized. As for the Co magnetic properties, the analyses of many experiments and the theoretical calculations have considered that the Co magnetism is uniform, while the $^{59}$Co NMR data indicate that a large interplay occurs between the atomic arrangements of Na ions and electronic properties of the CoO$_{2}$ planes (Mukhamedshin {\it et al.} 2014). The $^{59}$Co NMR permitted to evidence that some Co sites are non magnetic Co$^{3+}$ ions for which the six on site electrons fill the lower energy $t_{2g}$ sub-levels.

The concentration of such Co$^{3+}$ sites evolves markedly on the phase
diagram from $x=1$ to $x=1/2$. All Co sites are Co$^{3+}$ in Na$_{1}$CoO$_{2}$ which is a band insulator. For $x<1$ only a limited number of Na compositions with specific Na orderings can be synthesized. With decreasing Na content, the compound becomes metallic, and the fraction of Co$^{3+}$ sites is found to decrease and to vanish for  $x\leq 0.5$. As shown herafter, the non magnetic Co$^{3+}$ sites are found to form ordered arrays in a metallic Co background responsible for the magnetic properties. Detailed studies of the $^{59}$Co NMR for the other sites have been achieved mostly for two concentrations $x=2/3$ and $x=10/13\simeq0.77$ which are at boundaries between specific behaviors in the low $T$ electronic properties. On the $x=2/3$ phase NMR/NQR experiments have allowed the determination of both the 2D Na ordered structure and the 3D stacking of Na/Co planes. The Co$^{3+}$ ions are stabilized on 25\% of the cobalt sites Co1 arranged in a triangular sublattice. The holes are delocalized on the 75\% complementary cobalt sites Co2 which display a planar cobalt kagomé structure (as shown in ({\bf Fig.~\ref{Fig14}(c)}), with an average charge state close to Co$^{3.5+}$. The parameters of the Zeeman and quadrupolar Hamiltonians have been determined for all cobalt sites in the unit cell and this has permitted to evidence that the magnetic properties are dominated by the large $T$ dependence of the susceptibility at the Co2 sites ({\bf Fig.~\ref{Fig14}(b)}). They unexpectedly display a strong in-plane electronic anisotropy which could agree with an orbital ordering along the kagome sublattice organization. These experimental results resolve a puzzling issue by precluding localized moment pictures for the magnetic properties of sodium cobaltates. Indeed some theoretical speculations suggested that Na order pins local magnetic moments in a metallic and less magnetic bath.

Using spin lattice realaxation determinations of the experimental Korringa constant along the lines sketched in {\bf NMREPS} has permitted to establish that the electronic correlations within the metallic state do display an AF character for $x<2/3$ and become mostly ferromagnetic within the Co planes for$\ x\geq 2/3$  (Lang {\it et al.} 2008). These metallic ferromagnetic correlations do not yield any static magnetic order down to $T=0 $ for $2/3\leq x<0.75$.
For $x\geq 0.77$ ordered magnetism occurs at low $T$. The ferromagnetic Co layers display then 3D AF stackings .

In cuprates no charge order has been initially observed, the dopants being usually disordered. However recently specific observations have been done in samples in which ordered dopant states have been synthesized. This has been thoroughly studied in the case of YBCO, in which samples with well ordered O in the intermediate CuO$_{x}$ layers
have been produced. In such cases quantum oscillations due to Fermi
surface reconstructions at low $T$ have been observed. NMR experiments have also allowed to detect quadrupole effects on the $^{63}$Cu. Those indicate that the Fermi surface reconstructions are seen only in the underdoped compounds and the associated CDW order appears mostly when some dopant order is present (Wu {\it et al.} 2011). Those experiments indicate that the charge ordered state seems to compete with SC and sets in at temperatures significantly lower than the pseudogap $T^{\ast }$. Those observation have triggered a full set of new ideas about the occurrence of the pseudogap in the cuprates, but the controversies raised by these experiments are not fully resolved at this stage. While the pseudogap $T^{\ast }$ is apparently generic, the CDW order and its symmetry appear somewhat dependent of the cuprate family.\\

{\bf The nuclear quadrupole effects detected in NMR or NQR experiments usually permit to distinguish the charge environment of given nuclei. In correlated electronic solids this gives an access to charge differentiation on atomic sites or to charge density waves due to Fermi surface reconstruction when they occur. This has been illustrated here in the case of layered Na cobaltates or for the CDW which occurs well below the pseudogap $T^{\ast }$ in underdoped cuprates.}

\section{Insulating Magnetic states and spin liquids}
\label{Insulating Magnetic states and spin liquids}
We shall consider now non metallic cases where the strength of
electron correlations is large enough to fix the number of
electrons per site, which in the half filled cases corresponds to the Mott Insulating state. In those cases local moment occur on all sites even at the highest temperatures. Those moments interact with each-other through exchange interactions which are short range and can be in first
approximation treated as Heisenberg couplings
\[
\label{Exchange nn}H_{ij}=J_{ij}\ \ S_{i}\ .S_{j\ }.\]
Such couplings favor a ferromagnetic orientation of the spins for $J_{ij}<0$ and antiparallel orientations when $J_{ij}>0$. The actual ground states of such spin systems is therefore highly influenced by the geometrical arrangements of the spins on the underlying lattice. We shall therefore distinguish cases where the spins are ordered on a lattice, from situations where disorder effects prevail and dominate the spin properties.

\subsection{Ordered insulating magnetic states}
\label{Ordered insulating magnetic states}

If one considers only first nearest neighbor interactions, ferromagnetic
states always result for $J_{ij}<0\ $ while the spin order depends on the lattice if $J_{ij}>0$. In this latter case, if the lattice is bipartite, a N\'eel AF state with two sublattices with opposite spin directions is stabilized as it satisfies all pair interactions. Ferromagnetic (or ferrimagnetic if the spins on the two sublattices do not bear the same moment) states are usually easy to detect below their Curie temperature using macroscopic magnetization measurements. For antiferromagnetic compounds, the magnetic susceptibilities being low, macroscopic magnetic data do not always permit to determine whether a material is an antiferromagnet. In contrast, NMR measurements generally allow one to reveal the existence of frozen magnetism whatever the order of the magnetic state. The on site hyperfine couplings being quite large on nuclear spins of magnetic atoms, the internal field which appears on these sites at the ordering temperature often induces a loss of the NMR signal through the magnetic transition. The wipe out of part of the NMR signal is therefore a good signature of the occurrence of a magnetic state. 
\begin{figure}
\centering
\includegraphics[height=6.5cm,width=10cm]{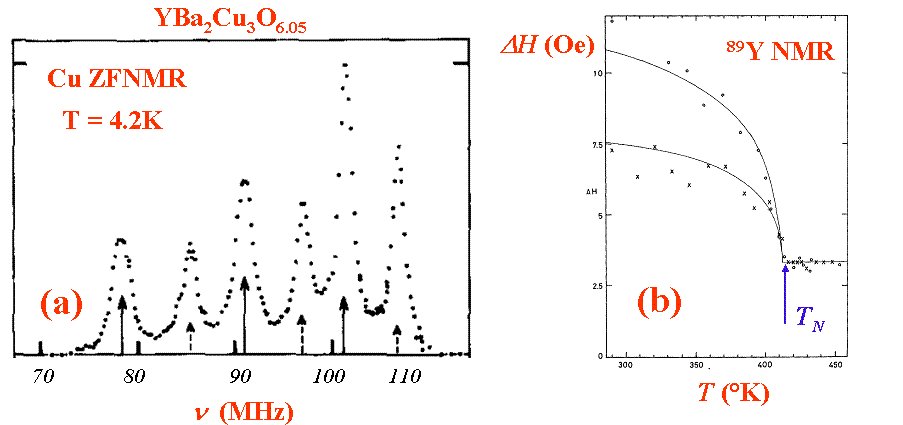}
\caption {The parent cuprate compound YBaCuO$_{6}$ is obtained by taking off the O atoms of the intermediate Cu layer in the structure of {\bf Fig.~\ref{Fig4}(a)}.{\bf(a)} NMR spectra of $^{63}$Cu and $^{65}$Cu taken at 4.2K in the AF state without applied field (Mendels and Alloul, 1988). Each nuclear spin species exhibits a three lines quadrupole splitting revealing that the internal field lays in the CuO$_{2}$ plane. {\bf(b)} The $^{89}$Y  NMR spectra obtained in an applied field has a width which increases sharply below $T_{N}$=410 K (Ohno {\it et al.} 1990).}
\label{Fig15}
\end{figure}
More of course can be learnt if the magnetic order is simple enough to
induce a well defined internal field on the magnetic sites, so that the
associated Larmor frequency might be detected in zero applied field. Simple examples can be taken from the undoped parent AF states of the cuprates, such as La$_{2}$CuO$_{4}$ or YBa$_{2}$Cu$_{3}$O$_{6}$. In the latter the Cu atoms of the CuO$_{2}$ planes are in a configuration $3d^{9}$ which bears a $S=1/2$ moment. Their $^{63}$Cu NMR can therefore be detected in the absence of any external applied field (see {\bf Fig.~\ref{Fig15}(a)}). Let us point out that the internal field being a measure of the local magnetization, can be taken as the order parameter of the magnetic phase, which vanishes above the N\`eel temperature. 
But usually the magnetic moment and the internal field associated with a
magnetic state are so large that the NMR of the magnetic site, or its
temperature dependence, might be hard to detect for various technical
reasons, the least being the fact that one usually has to perform a blind search of the Larmor frequency. In many cases the occurrence of a magnetic state is better detected using the NMR of near neighbor sites through weaker transferred hyperfine couplings. The NMR of such sites are detected in the paramagnetic state in a an applied field and, the induced internal fields being small, they can be seen to appear at the magnetic transition temperature. This has been done for instance using the La NQR spectra in La $_{2}$CuO$_{4}$. The use of the NMR of non magnetic sites of the structure adds a further local sensitivity to the actual arrangement of the moments in the magnetic phase. For instance, the induced internal field cancels on sites which are in a symmetric position within a bipartite AF structure. This occurs for the $^{63}$Cu site of the intermediate layer of Cu sites located between the CuO$_{2}$ planes in YBa$_{2}$Cu$_{3}$O$_{6}$. Being in a $3d^{10}$ non magnetic configuration the $^{63}$Cu NMR signal of those sites can only be detected with an applied field. Similarly the local field on the $^{89}$Y site which is located between the two AF CuO$_{2}$ layers
cancels in this compound. This does not forbid fully the possibility to
detect the occurrence of AF on those sites as the unavoidable defects in the crystalline structure does yield a small imbalance of the local field which is sufficient to be seen as a broadening of the $^{89}$Y NMR signal at $T_{N}$(see {\bf Fig.~\ref{Fig15}(b)}).

The physical properties of ordered magnetic states are quite dependent of the actual atomic ordering of the moments into spin chains, ladders or even more complicated structures. A simple review of the diverse types of low dimensional magnetism would require a full independent article to be written. Let us however mention the specific case of SrCu$_{2}$(BO$_{3}$)$_{2}$, for which NMR data has been quite useful. In the CuBO$_{3}$ planes of this compound, the Cu$^{2+}$ ions ($S=1/2$) form dimers and therefore the zero field ground state is a collection of singlets. In sufficiently large applied field ($H_{0} >20$ Tesla) triplet excitations of the dimers can be stabilized and magnetization plateaus are observed for well defined field values. $^{63}$Cu and $^{11}$B  NMR experiments taken in the very high fields corresponding to these plateaus give a beautiful detailed evidence for super-cell magnetic orderings of the triplets (Takigawa {\it et al.} 2004).
    
Through those simple examples we have shown that NMR techniques permits one to detect the occurrence of magnetic states, and this can be an important approach for some specific problems and to unravel cases on powder samples. But of course neutron scattering is certainly a privileged technique to study both the static magnetic order and the magnon excitations in magnetic solids as soon as single crystals are available. We have however shown cases where NMR permits to achieve observations not accessible to neutron scattering experiments, such as the difference between magnetic sites in the high field measurements in SrCu$_{2}$(BO$_{3}$)$_{2}$. Similarly NMR did permit ({\bf section~\ref{Charge differentiation and orbital order}}) to evidence in Na$_{x}$CoO$_{2}$ a  charge disproportionation of the Co sites in the paramagnetic state. This disproportionation also occurs for the compositions $x\geqslant 0.77$ which display a magnetic order at low $T$, but was not seen in neutron scattering data. It is however reflected by the large difference of local fields detected on the distinct Co sites in zero field NMR experiments in the AF state.

\subsection{Disordered magnetism and spin glasses}
\label{Disordered magnetism and spin glasses}
If, due to some disorder effects, both ferro and antiferromagnetic
couplings occur in a magnetic material, a frozen magnetic state might occur at low $T$, though without any specific magnetic order and any resultant macroscopic magnetization. This situation has been initially promoted in the case of metallic $sp$ systems with a large concentration of substituted magnetic impurities. As can be anticipated from the spin polarization studies illustrated in {\bf NMREPS}, the impurity local moments are coupled by RKKY Heisenberg interactions, the $J_{ij} $ being then of either sign depending on the relative positions of the impurities in the lattice. In this situation a random orientation of frozen spins prevails below a temperature $T_{g}$, naturally qualified then as the spin glass freezing temperature. The latter cannot be considered as a phase transition temperature and depends somewhat on the time window of the probe used to evidence the slowing down of the magnetic fluctuations. However an important signature of the occurrence of such a state is usually a sharp peak of the spin susceptibility detected at $T_{g}$. In NMR experiments $T_{g}$ can be detected by a loss of intensity of the NMR of the magnetic sites, or by a broadening of the NMR of the non magnetic sites probes. Apart the fact that the transition temperature is not as well defined as for an ordered magnetic state, the largest differences with ordered cases which occurs in these spin glass systems can be found in the spin dynamics. In ordered spin structures, long wavelength exitations (called magnons) can be defined at low $T$ and govern the reduction of the local magnetization
with increasing $T$, that is the decrease of the order parameter in the
ordered spin states. In spin glass phases, low energy magnetic excitations on a much shorter length-scale may dominate and drive fully the spectrum of magnetic excitations. Such differences have been exemplified by nuclear spin $T_{1}$ data taken at low $T$.

\subsection{Frustrated quantum states and spin liquids}
\label{Frustrated quantum states and spin liquids}
\begin{figure}
\centering
\includegraphics[height=9cm,width=12cm]{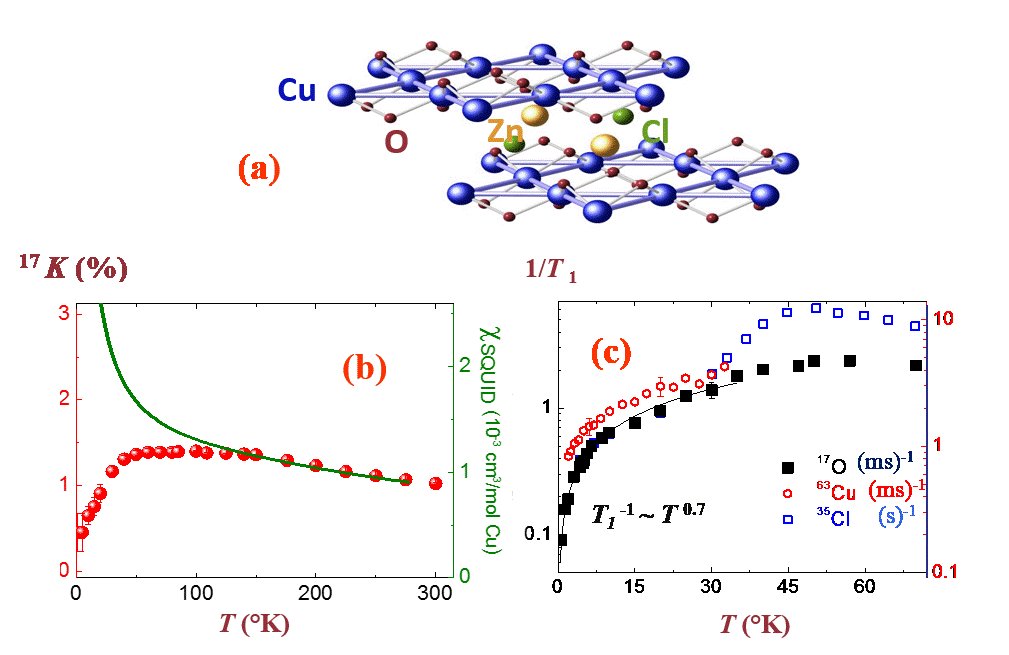}
\caption{{\bf(a)} Atomic structure of the Herbertsmithite. Here the Cu form Kagome layers separated by Zn layers. {\bf(b)} The spin susceptibility measured by $^{17}$O NMR shift data increases with decreasing $T$ but vanishes when $T$ approaches $0$ K. {\bf(c)} The spin lattice relaxation rate $1/T_{1}$ of all nuclear spins in the structure display a power law $T$ dependence at low $T$  (from Olariu {\it et al.} 2008).}
\label{Fig16}
\end{figure}
A rather interesting novel situation has been promoted in the last decade by the study of specific lattices in which frustration prohibits the formation of ordered spin structures in the absence of any disorder. By frustration we mean that the geometry of the lattice, or the spatial dependence of the interaction between sites, are such that all pair-wise interactions between sites cannot be minimized altogether. The frustration effects are favored if one organizes crystal structures by edge or corner sharing of triangles. They already occur in the 2D triangular lattice, in which ordered states can only occur with neighboring spins with relative orientations at $120\deg$.
One of the most famous regular case which has been studied at length
recently is that of the 2D Kagome structure, in which a macroscopic
degeneracy of the ground state could be expected. If furthermore the spins involved in such a structure are $S=1/2$, the quantum fluctuations are expected to be large and an ordered state is not anticipated even at $T=0$. The resulting spin state is considered as a {\bf spin liquid}. The archetype of such a Kagome spin 1/2 compound has been found in a
natural mineral, the Herbertsmithite, which could be synthesized after its discovery ({\bf Fig.~\ref{Fig16}(a)}). In that case the spin susceptibility, as measured by the $^{17}$O NMR shift in powder samples is still paramagnetic at $T<<J/k_{B}$, where $J\simeq 200 $ K. It is  even seen to decrease at low $T$ and frozen spin order is not detected down to 100 mK (Olariu {\it et al} 2008). This is corroborated by the
persistence of magnetic fluctuations down to the lowest $T$ as monitored by the power law increase of the spin lattice relaxation $1/T_{1}$ data ({\bf Fig.~\ref{Fig16}(b)}) (Carretta and Keren 2011). 
These experiments therefore highlight the importance of NMR in such systems, which permit to isolate the susceptibility of the Kagome plane and give direct evidence for the  dynamical character down to the lowest $T$. Such spin liquid states are viewed as ideal realizations of quantum states in which the spins are paired in $S=0$ dynamic resonating valence bond states ({\bf RVB}). Such states were initially suggested by P.W. Anderson (1987) to describe the parent state of the cuprates, a proposal which triggered an intense and still on-going activity on geometrically frustrated antiferromagnets. The RVB suggestion did not turn out to be true for he cuprates, although various authors consider that at low doping such RVB states are restored, hence the loss of AF state.\\

{\bf In insulating magnetic compounds, NMR experiments permit to differentiate ordered spin states from cases such as spin glasses and spin liquids for which disorder or frustration effects dominate. Disorder in the magnetic interactions generates spin glass frozen states. When frustration effects dominate, as in Kagome or similar frustrated lattices, frozen spin states are inhibited by quantum fluctuations and spin liquid states may survive down to T=0. Those represent the best realization of Resonating Valence Bond states (RVB).}

\section{Metallic magnetic states and Mott insulator to metal transition}
\label{Metallic magnetic states and Mott insulator to metal transition}
So far we have considered either Mott insulator local moment systems or
metallic states generated by doping a Mott insulator, as for the cuprates. But most of the intriguing correlated electron systems revealed in the last years are cases in which magnetism and SC coexist. We shall consider such cases in {\bf section~\ref{Magnetism and Superconductivity}}. But the case which has been considered for long as the simplest illustration of the correlated electron states is that of a Mott insulator which is driven into a metallic state by an applied pressure. This Mott transition should indeed exemplify the change of electronic properties induced by an increase of the $t/U$ ratio. However few compounds do exhibit a metal insulator transition which could be considered as a Mott transition. The most investigated compound, which has been considered as a prototypical case since the 1960's, is the vanadium oxide V$_{2}$O$_{3}$, which undergoes an insulator to metal transition with increasing pressure or temperature above room $T$. Although this metal insulator transition is always taken as an example, it has been mostly studied in samples with slight Cr or Ti substitution on the V site that induces some doping and an uncontrolled amount of disorder. For many other cases studied, a change in the atomic structure of the compound usually occurs at the metal insulator transition, as for instance in doped VO$_{2}$,
so that band effects interfere with electronic correlations. Most of these compounds have been studied dominantly by transport and optical
experiments. Here we shall focus in {\bf section~\ref{Organic and fulleride compounds}} on some new organic and fulleride compounds in which a Mott transition towards a SC state could be evidenced recently using dominantly NMR studies of the magnetic properties. 
\subsection{Magnetism and Superconductivity}
\label{Magnetism and Superconductivity}
While magnetism was considered to be incompatible with SC, searches for
compounds where these two behaviors would coexist have been very active in the 1970s. And indeed in many actual compounds with poly-atomic unit cells, magnetism can be sustained by Mott localized states on some specific atoms of the unit cell. Those may have very little overlap with the extended electron states involved in bands constructed from distinct atomic orbitals in the unit cell. Examples can be found in some compounds involving rare earth atoms and lighter elements such as the family of compounds discovered by Chevrel (see a review by Fischer 1978), an archetype being HoMo$_{6}$S$_{8}$. Here the Mo$_{6}$S$_{8}$
units give rise to $4d$ metallic bands while the rare earth Ho$^{3+}$ order ferromagnetically below $T_{C}=0.69$ K, so that magnetism and metallicity can be considered as independent. In that case a classic BCS
superconductivity with $T_{c}=1.2$ K is sustained by the $4d$ bands. This $T_{c}$ being small, superconductivity is nevertheless suppressed below $T_{C}=0.69$ K by the large internal field induced on the Mo sites in the magnetic ordered state of Ho$^{3+}$. Another example is again that of the YBa$_{2}$Cu$_{3}$O$_{6+x}$ cuprate family in which Y can be replaced by any rare earth $4f$ element. For instance in the compound GdBa$_{2}$Cu$_{3}$O$_{6+x}$, the Gd$^{3+}$ ions remain well isolated with a local magnetic moment of $7\mu_{B}$ very weakly coupled to the electronic states of the CuO$_{2}$ plane which are responsible for the 2D correlated metallic behavior and superconductivity. The coupling is so small that the metallic and SC properties of the CuO$_{2}$ planes are absolutely identical to those of YBa$_{2}$Cu$_{3}$O$_{6+x}$ with $
T_{c}$ values which differ by at most 1 K. However the small transfer
integrals are sufficient to induce a weak coupling between Gd ionic spin
states with an antiferromagnetic ordering of the Gd moments below a Néel
temperature $T_{N}=2.25$ K.

Such specific situations of coexistence between magnetism and metallic behavior are not so difficult to understand and to model. In cuprates, from what we have seen in the previous chapters, the strange metal occurs in the correlated band of the CuO$_{2}$ planes, which remains the
same for the Y and Gd compounds. But the actual situations which of
course trigger most interest are those where correlations and metallic transport occur on electronic bands of the same atomic species as in the CuO$_{2}$ plane of the cuprates or in some of the the Heavy Fermion compounds introduced in {\bf section \ref{Heavy Fermions and Kondo Lattices}}. In such compounds the metallic behavior can be
induced either by doping as for the cuprates, or by pressure, a situation we shall consider in more detail below for organic compounds.
\subsection{Organic and fulleride compounds}
\label{Organic and fulleride compounds}
Recently, metal insulator transitions have been found to occur under applied pressure in organic materials either of the BEDT-TTF family or in the fulleride compound Cs$_{3}$C$_{60}$. The transfer integrals between the organic or the fullerene C$_{60}$ molecules in such compounds are mostly occurring through Van der Waals interactions. These materials are therefore highly compressible and rather weak applied pressures are sufficient to control the bandwidth so that a large range of $t/U$ values can be spanned in their phase diagrams. For instance a transition from a 2D SDW metal to a superconductor can be seen in one of the (BEDT-TTF)$_{2}$X compounds to be discussed below. 

\begin{figure}
\centering
\includegraphics[height=7cm,width=12cm]{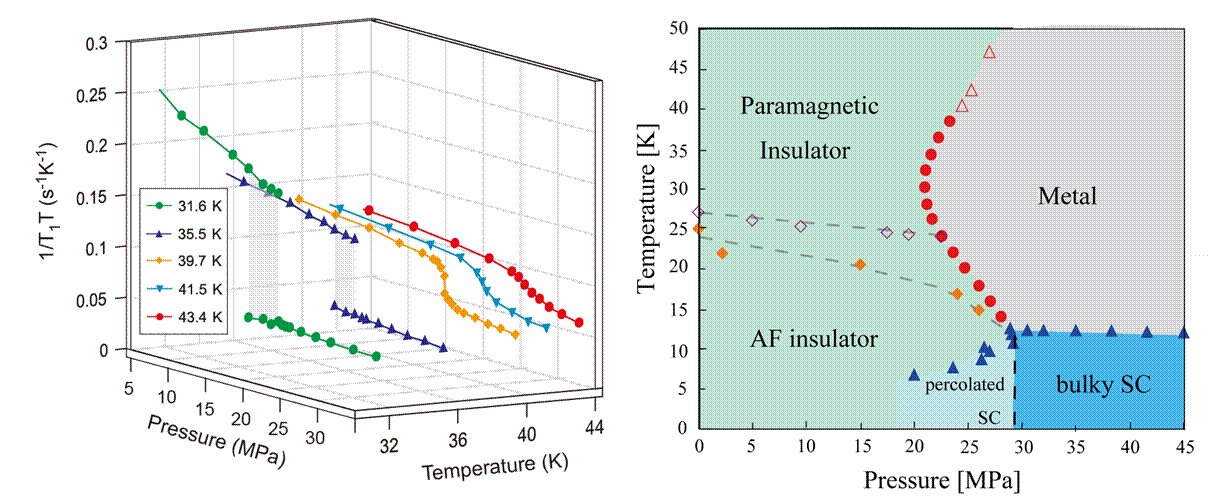}
\caption{{\bf(a)} The $(T_{1}T)^{-1}$ data taken at fixed $T$ versus pressure in the organic (BEDT-TTF)$_{2}$X compounds evidence a first order transition from the Mott insulating to the metallic state. {\bf(b)} The pressure of the Mott transition shifts with increasing $T$ and defines the red transition line. The latter crosses over towards a continuous transition above a critical temperature. The  magnetic and SC regimes are delineated in the phase diagram (from Kagawa {\it et al.}, 2009).}
\label{Fig17}
\end{figure}

As for Cs$_{3}$C$_{60}$, which is a 3D Mott insulator at ambient pressure, it is driven into a correlated SC state by application of a few kbar pressure. For both types of materials, the NMR has been quite useful in the determination of their phase diagrams. Being sensitive to the magnetic properties, NMR data have also permitted thorough
characterization of the magnetic properties on the insulating side of the Mott transition.
\subsubsection{Structures and phase diagrams}
\label{Structures and phase diagrams}
In the (BEDT-TTF)$_{2}$X compounds the organic molecules have a quasi
triangular structure and the phase diagram depends somewhat of the actual anion X introduced between the organic salt layers. NMR permitted to evidence that in the Mott Insulator case the compound with X=Cu$\left[ \text{N(CN)}_{2}\right] $Cl displays an AF state at low $T$ and becomes under pressure a SC with the highest $T_{c}=12$ K found so far in organic compounds ({\bf Fig.~\ref{Fig17}(b)})(Kagawa {\it et al.} 2009). For X=Cu$_{2}$(CN)$_{3}$ a less ordered magnetic phase not far
from a spin liquid occurs and the SC state $T_{c}=2.5$ K is lower (Kurosaki {\it et al.} 2005). This has been ascribed to changes in the anisotropy ratio $t/t'$ of transfer integrals in the distorted triangular lattice, which corresponds to an increase of the frustration in the latter compound. This system therefore exhibits a unique situation of a pressure induced Mott transition between a frustrated spin liquid and a SC state. 

Although the fullerene compound has a 3D cubic structure, quite remarkable analogies have been found as two distinct phases of Cs$_{3}$C$_{60}$ can be synthesized as has been evidenced from their $^{133}$Cs NMR spectra and powder X-ray diffraction data. Here-again a well defined Néel AF order is found to set in at $T_{N}=47$ K in the insulating phase of the A15 structure ({\bf Fig.~\ref{Fig18}(a)}). On the contrary in the fcc-Cs$_{3}$C$_{60}$ phase, which is not
a bipartite structure, more frustration occurs and spin liquid like magnetic fluctuations dominate down to much lower temperatures in the Mott phase (Ihara {\it et al.} 2010a). A still poorly characterized frozen spin state only occurs below a few K in this fcc-Cs$_{3}$C$_{60}$ phase. Finally, in these Cs$_{3}$C$_{60}$ compounds the lattice structure is not found to be modified at the Mott transition, both from x- ray data and from the NMR spectra. In the fcc phase the three usually  detected Cs sites are seen both in the Mott insulating and metallic states. In the A15-Cs$_{3}$C$_{60}$, the Cs occupies a single crystallographic site which is not in a cubic environment.
Therefore a seven line quadrupole splitting $\nu _{Q}$ of the $^{133}$Cs
NMR ($I=7/2$) is detected. The analysis of the spectra through the
Insulator to Metal transition only reveals a variation of $\nu _{Q}$ though the transition which can be associated with a weak variation of the cell parameter.
Furthermore the $T=0$ first order Mott transition pressure is found to
evolve with increasing $T$ in all these compounds. In the organic salts the evolution of the $p(T)$ transition line stops at a critical point ($p_{c}$,$T_{MI}$), so that these Mott transitions have all the characteristics of a liquid vapor transition ({\bf Fig.~\ref{Fig17}(b)}) . Indeed at temperatures higher than $T_{MI}$ the compound can be brought continuously with increasing $p$ from the insulating paramagnetic state to the high $T$ bad metal regime without crossing any symmetry breaking line. This could be seen directly (Kagawa {\it et al.} 2009) in the
organic salts by $T_{1}$ data taken on $^{13}$C enriched samples, for
which the discontinuity at the transition disappears above the critical
point.

\subsubsection{SC properties near the Mott transition}
\label{SC properties near the Mott transition}
Concerning the SC properties the situation is quite distinct within these two types of compounds. In the organic salts the SC is suspected to be exotic with $d$ wave symmetry and to be mediated by AF fluctuations though very clear experiments establishing that are not yet available. 
\begin{figure}
\centering
\includegraphics[height=7cm,width=12cm]{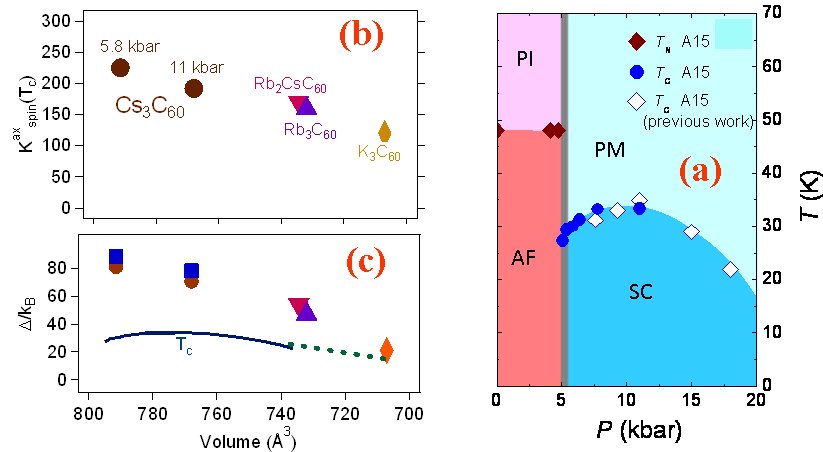}
\caption{{\bf(a)} Phase diagram of the {\it A} 15-Cs$_{3}$C$_{60}$ compound which is a Mott insulator at low $T$ with a N\`eel temperature independent of pressure up to the first order transition to the metallic and SC state at 5 kbar. The SC phase displays a dome of $T_{c}(p)$. {\bf(b)} The  $^{13}$C NMR shift anisotropy permits to monitor the variation of the normal state spin susceptibility. One clearly sees  that it increases with increasing the volume $V_{C60}$ per ball. {\bf (c)} The SC gap has been obtained from the $T$ variation below $T_{c}$ of $(T_{1}T)^{-1}$. One can see that it increases regularly with increasing $V_{C60}$, even when $T_{c}$ goes through its maximum value. Here the cases of the dense Rb$_{3}$C$_{60}$ and K$_{3}$C$_{60}$ compounds are used as  well (Wzietek {\it et al.} 2014).}
\label{Fig18}
\end{figure}
In the alkali fulleride $A_{3}$C$_{60}$ compounds the occurrence of $s$
wave superconductivity due to the interaction of electrons with the on ball optical phonon modes has been claimed quite early as a BCS picture (see {\bf RMNEPS}) with a gap $\Delta \simeq 1.75\ k_{B}T_{c}$ and a Hebel Slichter peak have been evidenced experimentally by NMR. However the occurrence of strong electronic correlations in the $A_{n}$C$_{60}$ compounds with $n\neq 3$ has been established and it has been claimed for long that these electronic correlations were essential as well to justify the high $T_{c}$ values found in the SC compounds. 

The discovery of the Cs$_{3}$C$_{60}$ expanded fulleride and of its Mott
transition is clearly confirming that idea and has permitted recently to
study the evolution of the SC state properties when approaching the Mott
transition. The phase diagram displays an original behaviour with a dome
shape of $T_{c}$ versus pressure which suggests a weakening of SC when
approaching the Mott transition. On the contrary $^{13}$C and $^{133}$Cs
NMR $T_{1}$ measurements of the superconducting gap $\Delta$ do give
evidence that $\Delta$ increases continuously when approaching the Mott
transition. Correspondingly, near the Mott transition $\Delta /k_{B}T_{c}$ increases regularly and exceeds the BCS value all over the $T_{c}$ dome ({\bf Fig.~\ref{Fig18}}).
So we can conclude that these organic and fullerene compounds are excellent examples permiiting to probe how electronic correlations near a Mott transition modify SC either mediated by phonons or by AF
fluctuations.\\

{\bf A limited number of compounds can be driven from a Mott insulating state to a superconducting state though a Mott transition by application of a hydrostatic pressure. We present here the two specific cases of 2D molecular organic triangular lattices and of 3D cubic alkali fullerides Cs$_{3}$C$_{60}$. In both cases NMR experiments permit to monitor the Mott transition from ordered AF states or a frustrated spin state towards superconductivity. This might permit to study continuously the evolution of the SC state with the increase of correlations near the first order Mott transition.}
\section{Summary}   
Most, if not all, the discoveries which have been done since the 1980's on correlated electron systems resulted from extensive experimental investigations. In this article I have shown that the NMR technique has been quite successful in this process. This has been exemplifed here by revealing some of the most important problems highly debated nowadays in correlated electron physics. The main impact of NMR comes about as preliminary experiments can be done on powder materials which are not perfect. The second aspect which highlights this technique is that NMR results, which are not surface sensitive, are quite reproducible so that most results presented here have usually been confirmed by independent investigations done in different laboratories on distinct sample materials. All this is eased by the fact that NMR experiments allow one to detect the incidence of defects and disorder effects on the very samples on which the data are taken. We have shown here that the introduction of specific defects, associated with the capability to detect locally their incidence is a powerful tool to unravel the properties of the pure material. A similar advantage has been  highlighted as well for the STM  techniques, which however require samples with sufficient surface quality.

As emphasized in many instances in this article, NMR permits to probe on the local scale a large set of relevant physical quantities ranging from the ordered spin or charge states, the spin fluctuations, the superconducting properties etc.. In insulating systems the NMR experiments in frustrated materials have permitted to establish the existence of spin liquid states which can be described by RVB physics. In metallic correlated electron systems, an important aspect is the ability to identify by NMR the electronic band(s) which are involved in the metallic state and to establish whether magnetism is associated with different degrees of freedom or due to the same bands. This permits for instance to delineate whether Kondo screened moments are at the origin of coherent bands in given Heavy Fermion systems. Similarly one can distinguish which sites or bands are responsible for magnetism and transport in charge ordered cases such as cobaltates. One would expect as well some band selectivity in multiband systems such as the newly discovered Fe pnictides or chalcogenides.

Concerning the metal insulator transitions, NMR helps to distinguish the situations where the transition is  linked with gap openings at the Fermi level due to CDW or Peierls transitions, from those initiated by disorder or by Mott physics. In fact the notion of Mott--Hubbard transition has revealed that it should be possible to treat the electronic structure of a Mott insulator and of a metallic state within a unique framework. Although major steps have been taken in this direction, this is far from being achieved. The developments currently under way on organic and fulleride compounds which undergo pressure induced Mott transitions toward metallic superconducting states should trigger theoretical developments at the forefront of the research on the physics of correlated electron systems.

Coming to SC, one of the points which has attracted most attention is its interference (destructive or constructive) with metallic magnetism. The cuprates are in that respect certainly exotic superconductors, in which the incidence of electron correlations and AF short range interactions can be essential to drive superconductivity, or at least enhance the SC transition temperatures. Many other materials have been shown to display situations where magnetism and SC are proximate to each-other in phase diagrams. In Fe superconductors (pnictides or chalcogenides) the phase diagrams are sometimes spanned by doping as in the cuprates, but in other families of compounds the phase diagrams are spanned by pressure control of the overlap integrals as for organic, heavy fermions or Cs$_3$C$_{60}$ compounds.

In most of these cases a thorough experimental characterization of the SC order parameter symmetry is needed prior to any determination of the pairing glue, and NMR data can be helpful in that respect. The d-wave symetry of the order parameter for cuprates is considered as the strongest indication that electronic correlations could be responsible for the pairing in these compounds. Even for materials less correlated than the cuprates the incidence of electronic correlations is definitely less detrimental to SC than initially expected and the existence of AF correlations in the material could as well be the boson field mediating the SC state. Many possible glues between electrons such as phonons, AF fluctuations, charge correlations near a Quantum Critical Point, have been considered and may be at work in distinct materials. But all this is far from being settled and requires thorough investigations specific to the various families of correlated eletcron materials.
\section{References}
\label{References}
\begin{enumerate}
\item Alloul H. (2012) ``From Friedel oscillations and Kondo effect to the pseudogap in cuprates”, published in a volume dedicated to the 90th anniversary of Jacques Friedel, in J Supercond. Nov. Mag. 25, 385, doi: \href{http://dx.DOI.org/10.1007/s10948-012-1472-x}{10.1007/s10948-012-1472-x}., arXiv:1204.3804.

\item Alloul H., Bobroff J., Gabay M. and Hirschfeld P. (2009) ``Defects in correlated metals and superconductors”, condmat/ 0711 0877, Review of Modern Physics  81, 45.

\item  Alloul H., Mahajan A., Casalta H. and Klein O., (1993) ``$^{89}Y$ NMR study of the anisotropy of the static and dynamic susceptibilities in YBa2Cu3O6+x"., Phys. Rev. Lett. 70, 1171.

\item  Alloul H., Mendels P., Casalta H., Marucco J.F. and Arabski J. (1991) ``Correlations between Magnetic and Superconducting Properties of Zn Substituted YBa2Cu3O6+x". Phys. Rev. Lett. 67, 3140. 

\item  Alloul H., Ohno T. and Mendels P. (1989) ``$^{89}Y$ NMR evidence for a Fermi liquid behaviour in YBa2Cu3O6+x "Phys.  Rev. Lett. 63, 1700  

\item Alloul H., Mendels P., Collin  G. and Monod P. (1988) ``$^{89}Y$ NMR study of the Pauli susceptibility of the CuO2 planes in YBa2Cu3O6+x"  Phys. Rev. Letters 61, 746. 

\item Alloul H., Hippert F. and Ishii H. (1979) ``NMR Evidence for Kondo deviations of the Mn Electron spin relaxation and g factor in CuMn", J. of Physics F - Metals, 9, 725.

\item Alloul H. (1977) ``Hyperfine studies of the static and dynamic susceptibilities of Kondo systems". Physica B 86-88, 449.

\item Alloul H. (1976) ``Satellite NMR study of the dynamic susceptibility of Fe in Cu above and below TK". J. de Physique Lettres 37, 205

\item Alloul H. (1975) ``Susceptibility and electron Spin lattice relaxation of Fe in Cu below TK a  63Cu NMR study"; Phys, Rev. Letters 35, 460.

\item  Anderson P. W., (1987) ``The resonating valence bond state in La2CuO4 and superconductivity" Science 235, 1196–1198.

\item  Asayama K., Zheng G.-Q., Kitaoka Y., Ishida K. and Fujiwara K., (1991) ``NMR study of high Tc superconductors", Physica C 178, 281 

\item Behnia K., Balicas L., Kang W., Jérome D., Carretta P., Fagot-Revurat Y., Berthier C., Horvatić M., Ségransan P., Hubert L., and Bourbonnais C., (1995) ``Confinement in Bechgaard Salts: Anomalous Magnetoresistance and Nuclear Relaxation", Phys. Rev. Lett. 74, 5272.

\item  Bobroff J., Mac Farlane A., Alloul H., Mendels P., Blanchard N., Collin G. and Marucco J. F. (1999) ``Spinless Impurities in High Tc Cuprates: Kondo Like Behaviour",Phys. Rev. Letters, 83, 4381.

\item  Bobroff J., Alloul  H., Mendels P., Viallet V., Marucco J. F. and Colson D.,  (1997) ``17O NMR evidence for a pseudogap in the monolayer HgBa2CuO$_4+\delta$ Phys. Rev. Lett. 78, 3757.

\item Carretta P and Keren A., (2011) ``NMR and muSR in highly frustrated magnets",page 79 in ``Springer series in solid state sciences 164", Editors  Lacroix C, Mendels P and Mila F

\item Curro N.J., (2009) ``Nuclear magnetic resonance in the heavy fermion superconductors", Rep. Prog. Phys. 72  026502, doi: \href{http://dx.DOI.org/10.1088/0034-4885/72/2/026502}{10.1088/0034-4885/72/2/026502}.

\item  Das J., Mahajan A. V., Bobroff J., Alloul H., Alet F. and Sorensen E. (2004)  “Comparison of S=0 and S=1/2 impurities in the Haldane chain compound Y2BaNiO5”, Phys. Rev. B 69, 144404 . 

\item Daybell M. D. and Steyert W. A., (1968) ``Localized Magnetic Impurity States In Metals: Some Experimental Relationships" Rev. Mod. Phys. 40, 380.

\item  Emery V. J. and Kivelson S. A., (1995) ``Importance of phase fluctuations in superconductors with small superfluid density" Nature 374, 434.

\item Fischer O., (1978) ``Chevrel phases: Superconducting and Normal state properties" Applied physics 16,1.

\item  Giamarchi T., (2004) ``Quantum Physics in one dimension" published by Oxford University Press. ISBN 0-19-852500-1.

\item Ihara Y., Alloul H., Wzietek P., Pontiroli D., Mazzani M. and  Ricc\`o  M., (2010a). “NMR study of the Mott transitions to superconductivity in the two Cs3C60 phases”,condmat/1003.2977, Phys. Rev. Lett. 104, 256402.

\item  Ihara Y., Wzietek P., Alloul H., Rümmeli M. H., Pichler Th. and Simon F., (2010b) ``Incidence of the Tomonaga-Luttinger liquid state on the NMR spin lattice relaxation in Carbon Nanotubes",  condmat /0910.1147, EPL 90, 17004.

\item  Ishida K., Mukuda H., Kitaoka Y., Mao Z. Q., Mori Y. and Maeno Y., (2000) ``Anisotropic Superconducting Gap in the Spin-Triplet Superconductor Sr2RuO4: Evidence from a Ru-NQR Study" Phys. Rev. Lett. 84, 5387. 

\item  Ishida K., Kawasaki Y., Tabuchi K., Kashima K., Kitaoka Y.,  Asayama K., Geibel C., and Steglich F., (1999) ``Evolution from Magnetism to Unconventional Superconductivity in a Series of CexCu2Si2 Compounds Probed by Cu NQR" PRL. 82, 5353. 

\item  Ishida K., Mukuda H., Kitaoka Y., Asayama K., Mao Z. Q., Mori Y. and Maeno Y., (1998) ``Spin-triplet superconductivity in Sr2RuO4 identified by 17O Knight shift" Nature 396 658--60. 

\item Janis V. and Augustinsky P., (2008) ``Kondo behavior in the asymmetric Anderson model: Analytic approach Phys Rev B 77, O85106

\item Julian S.R., Tautz F.S., McMillan G.J. and Lonzarich G.G.(1994)  ``Quantum oscillation measurements of band magnetism in UPt3 and CeRu2Si2" Physica B 199 - 200, 63 .

\item  Kagawa F., Miyagawa K. and Kanoda K., (2009) “Magnetic Mott criticality in a $\kappa$-type organic salt probed by NMR”, Nature Physics  5.

\item Kaminski A., Kondo T., Takeuchi T. and  Gu G., (2014) ``Pairing, pseudogap and Fermi arcs in cuprates", Philosophical Magazine, doi:  \href{http://dx.DOI.org/10.1080/14786435.2014.906758}{10.1080/14786435.2014.906758}.

\item Kitaoka Y.,  Ueda K., Kohara T., Kohori Y. and Asayama K.. (1987) in ``Nuclear Magnetic Resonance in Heavy Fermion systems: Theoretical and Experimental Aspects of Valence Fluctuations and Heavy Fermions", eds. L.C. Gupta and S.K. Malik (Plenum) 297.  

\item  Kuramoto Y. and Kitaoka Y.(2000) ``Dynamics of Heavy electrons", The International Series of Monographs in Physics, Oxford University Press,  (march 2000)     9780198517672.

\item  Kurosaki Y., Shimizu Y., Miyagawa K., Kanoda K. and Saito G., (2005) ``Mott Transition from a Spin Liquid to a Fermi Liquid in the Spin-Frustrated Organic Conductor $\kappa$-(ET)2Cu2(CN)3 Phys. Rev. Lett. 95, 177001.

\item  Lang  G., Bobroff J., Alloul H., Collin G. and Blanchard N., (2008) ”Spin correlations and cobalt charge states: Phase diagram of sodium cobaltates”, Physical Review B 78, 155116.

\item Leggett A.J., (1975) ``A theoretical description of the new phases of liquid 3He", Rev. Mod. Phys. 47, 331. 

\item  Mahajan A. V., Alloul H., Collin G. and Marucco J. F., (1994) ``$^{89}Y$ NMR Probe of Zn Induced Local Moments in YBa2(Cu1-yZny)3O6+x" Phys. Rev. Letters 72, 3100.

\item Matsuda K., Kohori Y., Koha T., (1996) ``Existence of Line Nodes in the Superconducting Energy Gap of Antiferromagnetic Superconductor URu 2Si${_2}$-$^{101}$Ru NQR Study-" J. Phys. Soc. Jpn. 65, 679. 

\item  Mendels  P. and Alloul H., (1988) ``Zero field NMR of the magnetic copper sites in antiferromagnetic YBa2Cu3O6+x ". Physica C 156, 355
\item  Mila F. and Rice M. (1989) ``Analysis of Magnetic Resonance Experiments in YBa2Cu307" Physica C157, 561.

\item  Millis A.J., Monien H. and Pines D., (1990) ``Phenomenological model of nuclear relaxation in the normal state of YBa2Cu3O7" Phys. Rev. B  42, 167.

\item Mukhamedshin I.R., Dooglav V, Krivenko S.A., Alloul H, (2014) `` Evolution of Co charge disproportionation with Na order in NaxCoO2 ''  Phys. Rev. B 90, 115151. 

\item  Mukhamedshin I.R. and Alloul H., (2011) “59Co NMR evidence for charge and orbital order in the kagomé like structure of Na2/3CoO2”, Phys. Rev. B 84, 155112. 

\item  Ohno T., Alloul H. and Mendels P., (1990) “89Y NMR Study of antiferromagnetic YBa2Cu3O6“ J. Phys. Society of Japan 59, 1139-1142.

\item   Olariu A., Mendels P.,  Bert F., Duc F., Trombe  J. C., de Vries M. A. and Harrison A., (2008) “17O NMR Study of the Intrinsic Magnetic Susceptibility and Spin Dynamics of the Quantum Kagome Antiferromagnet ZnCu3(OH)6Cl2”, Phys. Rev. Lett., 100, 087202. 

\item  Rullier-Albenque F., Alloul H. and Rikken G. (2011) “High Field Studies of Superconducting Fluctuations in High-Tc Cuprates: Evidence for a Small Gap distinct from the Large Pseudogap” Phys. Rev. B 84, 014522..

\item Singer P.M., Wzietek P., Alloul H., Simon F. and Kuzmany H., (2005) ``NMR Evidence for Gapped Spin Excitations in Metallic Carbon Nanotubes", cond-mat/0510195, Phys. Rev. Lett. 95, 236403.

\item  Takigawa M. Kodama K., Horvatic M, Berthter C.,Kageyama H., Ueda Y., Miyahara S., Becca F. and Mila F., (2004) ``The 18-magnetization plateau state in the 2D quantum antiferromagnet SrCu2(BO3)2: spin superstructure, phase transition, and spin dynamics studied by high-field NMR", Phys.B  346-347,27.

\item  Takigawa M., Reyes A. P., Hammel P. C., Thompson J. D., Heffner R. H., Fisk Z. and  Ott K. C., (1991) ``Cu and O NMR studies of the magnetic properties of YBa2Cu3O6.63 (Tc=62 K)"  Phys. Rev. B 43, 247.

\item  Takigawa M., Reyes A. P., Hammel P. C., Thompson J. D.,  Heffner R. H., Fisk Z. and Ott K. C., (1990) ``Cu and O NMR studies of the magnetic properties of YBa2Cu3O6.63 (Tc=62 K)"  Phys. Rev. B 43, 247.

\item Takigawa  M., Hammel P. C., Heffner R. H.. Fisk Z., 0tt K. C. and Thompson J. D., (1989) ``17O NMR Study of YBa2Cu307-y", Physica C 162-164 853.

\item  Teloldi F., Santachiara R. and Horvatic M.(1999) ``Y89 NMR Imaging of the Staggered Magnetization in the Doped Haldane Chain Y2BaNi1-xMgxO5", Phys. Rev. Lett. 83,412.

\item Timusk T., and Statt B., (1999) ``The pseudogap in high-temperature superconductors: an experimental survey", Rep. Prog. Phys. 62,61.

\item Walstedt R.E. (2008) ``The NMR probe of High Tc materials",  Springer tracts in Modern physics, 228, doi: \href{http://dx.DOI.org/10.10007/978-3-540-75565-4}{10.10007/978-3-540-75565-4}.
\item Wilson K. G. (1975) ``The renormalization group: Critical phenomena and the Kondo problem". Rev. Mod. Phys, 47, 773.

\item  Wu T., Mayaffre H., Kramer S., Horvatic M., Berthier C., Hardy W.N., Liang  R., Bonn D.A., and Julien M.-H., (2011) ``Magnetic-field-induced stripe order in the high temperature superconductor YBa2Cu3Oy"  Nature 477, 191.  

\item  Wzietek P.,  Mito T., Alloul H., Pontiroli D., Aramini A.and Ricc\`o, M., (2014) “NMR study of the Superconducting gap variation near the Mott transition in Cs3C60” arXiv:1310.5529, Phys. Rev. Lett.  112, 066401.

\item Wzietek P., Creuzet F., Bourbonnais C., Jerome D., Bechgaard K., Batail P., ( 1993) ``Nuclear relaxation and electronic correlations in quasi-one-dimensional organic conductors. II. Experiments" Journal de Physique I, 3 171. 

\item Yang Y.,  Fisk. Z., Lee H., Thompson J.D. and Pines D., (2008) ``Scaling the Kondo lattice", Nature 454, 611.

\item  Zhang F. C.and Rice T. M. (1988) ``Effective Hamiltonian for the superconducting Cu oxides", Phys. Rev. B 37, 3759.

\end{enumerate}

\section{Further reading} 
\begin{enumerate}

\item Abragam A., (1961) ``The Principles of nuclear magnetism" (Oxford: Clarendon Press, London).

\item Alloul H., (2011) ``Introduction to the physics of Electrons in Solids" Graduate texts in Physics,  Springer –Verlag (Heidelberg), ISBN 978-3-642-13564-4, doi:  \href{http://dx.DOI.org/10.1007/978-3-642-13565-1}{10.1007/978-3-642-13565-1}.

\item Ashcroft Neil W., Mermin N. David, (1976), ``Solid State Physics", Saunders College, ISBN 0030493463, 9780030493461.

\item Balents L., (2010),  ``Spin liquids in frustrated magnets",  NATURE, 464,11, doi:\href{http://dx.DOI.org/10.1038/nature08917}{10.1038/nature08917}.

\item Fischer K. H. and Hertz J. A., (1993), ``Spin Glasses" (Cambridge:Cambridge University Press).

\item Giamarchi T., (2003), ``Quantum Physics in one dimension" Oxford University Press,  ISBN 0-19-852500-1 (published Dec 2003 (UK); Feb 2004 (US)).

\item Hewson A.C., (1993), ``The Kondo Problem to Heavy Fermions” (Cambridge University Press). 

\item Jerome D. and Schulz H. J., (2002), ``Organic Conductors and Superconductors",  Adv. Phys., 51, 293.

\item Kittel C., (2005), ``Introduction to Solid State Physics", 8th Edition, Wiley, ISBN 047141526X, 9780471415268.

\item Lacroix C., Mendels P. and Mila F., (2011), ``Introduction to Frustrated Magnetism", Springer.

\item Mackenzie A. P. and Maeno Y., (2003),  ``The superconductivity of Sr2RuO4 and the physics of spin-triplet pairing" Rev. Mod. Phys. 75, 657 - 712 

\item Moriya T., (1985), ``Spin Fluctuations in Itinerant Electron Magnetism", Springer Series in Solid-State Sciences Volume 56 ISBN: 978-3-642-82501-9. 

\item Mott N., (1974), ``Metal Insulator transitions" (Taylor \& Francis ), ISBN 0850660793, 9780850660791.

\item Mydosh J. A., (1993), ``Spin Glasses: An Experimental Introduction", (London: Taylor and Francis).

\item Slichter C. P., (1963), ``Principles of Magnetic Resonance", Harper  and Row (Springer-Verlag, New York, (1989), 3rd ed.

\item Tinkham M., (1996), ``Introduction to Superconductivity", Dover Publications, ISBN 0486134725, 9780486134727.

\item Walstedt R.E., (2008), ``The NMR probe of High Tc materials", Springer tracts in modern physics, 228, doi: \href{http://dx.DOI.org/10.10007/978-3-540-75565-4}{10.10007/978-3-540-75565-4}.

\end{enumerate}

\section{See also}
\begin{enumerate}

\item ``Bardeen Cooper Schrieffer theory",  Leon Cooper and Dimitri Feldman (2009), Scholarpedia, 4(1):6439.

\item ``Goodenough-Kanamori rule",  John B. Goodenough (2008), Scholarpedia, 3(10):7382.

\item ``Kondo effect",  Alex C Hewson and Jun Kondo (2009), Scholarpedia, 4(3):7529.

\item ``Magnetism: mathematical aspects",  Vieri Mastropietro and Daniel C. Mattis (2010), Scholarpedia, 5(7):10316.

\item ``NMR studies of electronic properties of solids”, H. Alloul,  Scholarpedia,   9(9):32069. (2014), doi: \href{http://dx.DOI.org/10.4249/scholarpedia.32069}{10.4249/scholarpedia.32069}.

\item ``Strongly Correlated Electrons in Solids", H. Alloul, Scholarpedia,  99(7):32067.  (2014), doi: \href{http://dx.DOI.org/10.4249/scholarpedia.32067}{10.4249/scholarpedia.32067}.
\end{enumerate}
\end{document}